\documentclass[%
 reprint,
 longbibliography,
superscriptaddress,
 footinbib,
 amsmath,
 amssymb,
 aps,
 pra,
]{revtex4-2}

\usepackage[utf8]{inputenc}
\usepackage{graphicx}
\usepackage{dcolumn}
\usepackage{bm}
\usepackage[dvipsnames]{xcolor}
\usepackage[colorlinks=true,linkcolor=teal,citecolor=RubineRed,urlcolor=RubineRed]{hyperref}
\usepackage{physics}
\usepackage{siunitx}
\usepackage{float}
\usepackage{tikz}
\usetikzlibrary{quantikz}
\usepackage{mathbbol}
\usepackage{enumitem}
\usepackage{xfrac}
\usepackage{simpler-wick}

\newcommand{\qgate}[1]{\textsc{#1}}

\newcommand{\tavg}[1]{\overline{#1}}
\newcommand{\simhz}[1]{\SI{#1}{\MHz}}
\newcommand{\zpf}{\varphi_{\mathrm{zpf}}}
\newcommand{\dtwopi}{/ 2 \pi}
\newcommand{\secref}[1]{Section~\hyperref[#1]{\ref*{#1}}}
\newcommand{\appref}[1]{Appendix~\hyperref[#1]{\ref*{#1}}}
\newcommand{\tabref}[1]{Table~\hyperref[#1]{\ref*{#1}}}
\newcommand{\figref}[1]{Fig.~\hyperref[#1]{\ref*{#1}}}
\newcommand{\sfigref}[2]{Fig.~\hyperref[#1]{\ref*{#1}(#2)}}
\newcommand{\orcid}[1]{\href{https://orcid.org/#1}{\includegraphics[width=8pt]{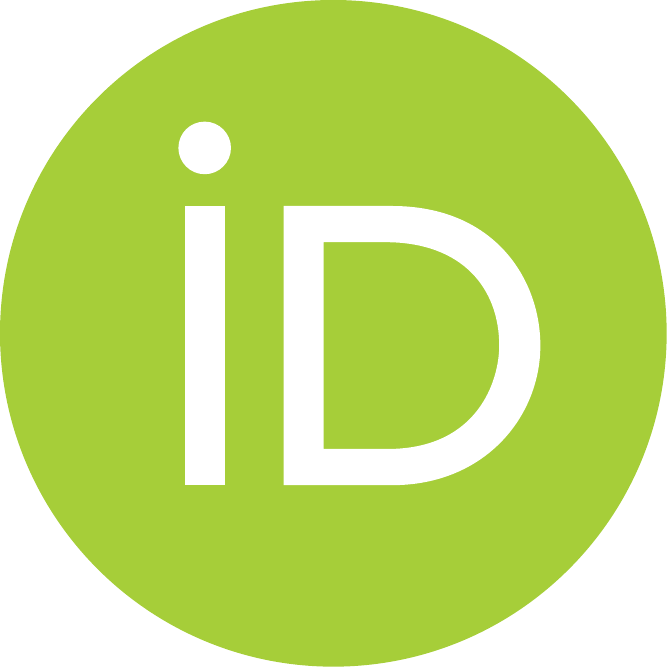}}}
\newcommand{\timo}[1]{{#1}}
\newcommand{\fernando}[1]{{#1}}

\begin{document}

\title{Designing Kerr Interactions 
for Quantum Information Processing \\
via Counterrotating Terms of Asymmetric Josephson-Junction Loops}

\author{Timo Hillmann\orcid{0000-0002-1476-0647}
}
\email{hillmann@chalmers.se}
\affiliation{Department of Microtechnology and Nanoscience (MC2), Chalmers University of Technology, SE-412 96 Gothenburg, Sweden}
\author{Fernando Quijandría\orcid{0000-0002-2355-0449}
}
\thanks{This work was done before F.Q. joined OIST.}
\affiliation{Quantum Machines Unit, Okinawa Institute of Science and Technology Graduate University, Onna-son, Okinawa 904-0495, Japan}


\begin{abstract}
    Continuous-variable systems realized in high-coherence microwave cavities are a promising platform for quantum information processing. 
    While strong dynamic nonlinear interactions are desired to implement fast and high-fidelity quantum operations, static cavity nonlinearities typically limit the performance of bosonic quantum error-correcting codes.
    \fernando{Here we study theoretical models of nonlinear oscillators describing superconducting quantum circuits with asymmetric Josephson-junctions loops. 
    Treating the nonlinearity as a perturbation, we derive effective Hamiltonians using the Schrieffer-Wolff transformation.
    We support our analytical results by numerical experiments and show that the effective Kerr-type couplings can be canceled by an interplay of higher-order nonlinearities.
    This can be better understood in a simplified model supporting only cubic and quartic nonlinearities. Our results show that a cubic interaction allows to increase the effective rates of both linear and nonlinear operations without an increase in the undesired anharmonicity of an oscillator which is crucial for many bosonic encodings.
    }
\end{abstract}
\maketitle

\fernando{
\section{Introduction}
It is still uncertain which physical platform will enable fault-tolerant quantum information processing.
Besides the traditional usage of two-level systems for fault-tolerant quantum computing (FTQC), for example realized in trapped ions~\cite{bruzewicz_trapped-ion_2019, wineland_nobel_2013}, superconducting circuits~\cite{devoret_superconducting_2013, blais_circuit_2021} and quantum dot spin systems~\cite{kloeffel_prospects_2013}, continuous-variable (CV) architectures with infinite-dimensional Hilbert spaces have emerged as a promising path to reduce the required hardware overhead~\cite{ofek_extending_2016, hu_quantum_2019, gertler_protecting_2021}.
In these architectures the underlying hardware consists of quantized radiation, either in optical devices~\cite{haroche_nobel_2013}, optomechanical cavities~\cite{aspelmeyer_cavity_2014} or superconducting microwave cavities~\cite{joshi_quantum_2020}.
In particular, the latter have seen an increased interest in the last decade due to their good coherence properties~\cite{romanenko_three-dimensional_2020, kudra_high_2020, heidler_observing_2021} and ease of integration with other circuit QED (cQED) elements such as Josephson junctions (JJ)~\cite{axline_architecture_2016}.
These nonlinear, low-loss, inductive elements enable nonlinear photon interactions 
required for universal quantum computation with CV systems~\cite{lloyd_quantum_1999} .
However, JJ's also imprint a finite (generalized) Kerr anharmonicity onto the oscillator spectrum which is undesired in many applications as it limits the fidelity of bosonic operations as well as the lifetime of encoded information~\cite{campagne-ibarcq_quantum_2020, gertler_protecting_2021, ma_error-transparent_2020}.
Josephson-junction-based devices can also be used to implement (nonlinear) interactions between two or more bosonic modes~\cite{pfaff_controlled_2017, lescanne_irreversible_2020}.

The rotating wave approximation (RWA) is commonly employed to derive effective non-linear Hamiltonians in cavity QED setups. The RWA gives an accurate description of the slow dynamics of a system by neglecting contributions from fast oscillating terms (compared to the natural dynamics of the unperturbed system). A natural scenario in which corrections to the RWA become necessary corresponds to driven systems. By considering linear resonators dispersively coupled to a transmon qubit, recent works have studied the effective non-linear Hamiltonians which result from the off-resonant drive of the transmon by a single or multiple tones within and beyond the number-split regime~\cite{wang_photon-number-dependent_2021,zhang_drive-induced_2021}. In addition, the effects of strong drives in the design of parametric gates realising two-qubit interactions have also been studied~\cite{petrescu_accurate_2021-1}. Drive-induced effects can be described either by means of time-dependent perturbation theory or Floquet theory~\cite{venkatraman_static_2021}.

A different scenario and the focus of this work are microwave circuits in which the nonlinearity is provided by generalized asymmetric Superconducting Quantum Interference Devices (SQUIDs) such as the 
Superconducting Nonlinear Asymmetric eLement (SNAIL) and the Asymmetric Threaded SQUID (ATS)~\cite{frattini_3-wave_2017, lescanne_exponential_2020}.
These correspond to Josephson junctions arranged in a loop configuration where the asymmetry refers to the difference in the Josephson energies of the participating junctions.
In these circuits, non-linear photon-photon interactions of different orders are available beyond even-order interactions as it is the case of JJs or symmetric SQUIDs. 
These novel circuits gained relevance in the context of quantum-limited amplification as they realize a native third-order nonlinear photon interaction with the possibility of tuning the fourth-order nonlinear coupling to zero when biased with an appropriate external magnetic flux. 
In the absence of time-dependent driving fields, the resulting non-linearities are diagonal or equivalently, correspond to terms containing an equal number of creation and annihilation operators (Kerr-type nonlinearities). Renormalization of these is possible through higher-order photon-photon interactions that are off-resonantly enabled. Thus, a careful study of these perturbative corrections will enable 
the design of experiments in which non-linear effects in a non-driven system are minimized. 

Leading order corrections to the RWA can be studied by using the James’ effective Hamiltonian method or the (time-dependent) Schrieffer-Wolff (SW) transformation. Interestingly, we show that incorporating James’ method into the SW allows to naturally derive an iterative relation to construct the generators of the SW transformation up to an arbitrary perturbative order which, to the best of our knowledge, was missing in the literature. This relation is general and does not restrict to the systems studied in this work.
The validity of our method is tested through numerical experiments on a simplified model of a SNAIL/ATS resonator. Our results
confirm
that it is possible to significantly reduce spurious Kerr-type interactions through the interplay of different order nonlinearities.

The subsequent sections are structured as follows.
We begin in \secref{sec:motivation_hamiltonian} by motivating the choice of the analyzed system Hamiltonians and give examples for Josephson-junction-based devices that readily implement the desired interactions with tunable couplings.
In \secref{sec:deriving_eff_hamiltonians} we briefly review 
the James' and the SW transformation methods
to obtain an effective Hamiltonian. Here we derive 
the relation to construct the generators of the SW transformation. 
\secref{sec:analytical_results} contains the main results of this work, that is, perturbative analytical expressions for the self-Kerr coupling for Hamiltonian with higher-order nonlinearities.
We support our analytical results numerically in \secref{sec:numerical_results} through exact diagonalization techniques, observation of reduced ``scrambling" of quantum information and improved state preparation fidelities.
Lastly, in \secref{sec:discussion_conclusion}, we summarize our results and discuss current and future implications as well as open questions.}
\begin{figure}
    \centering
    \includegraphics{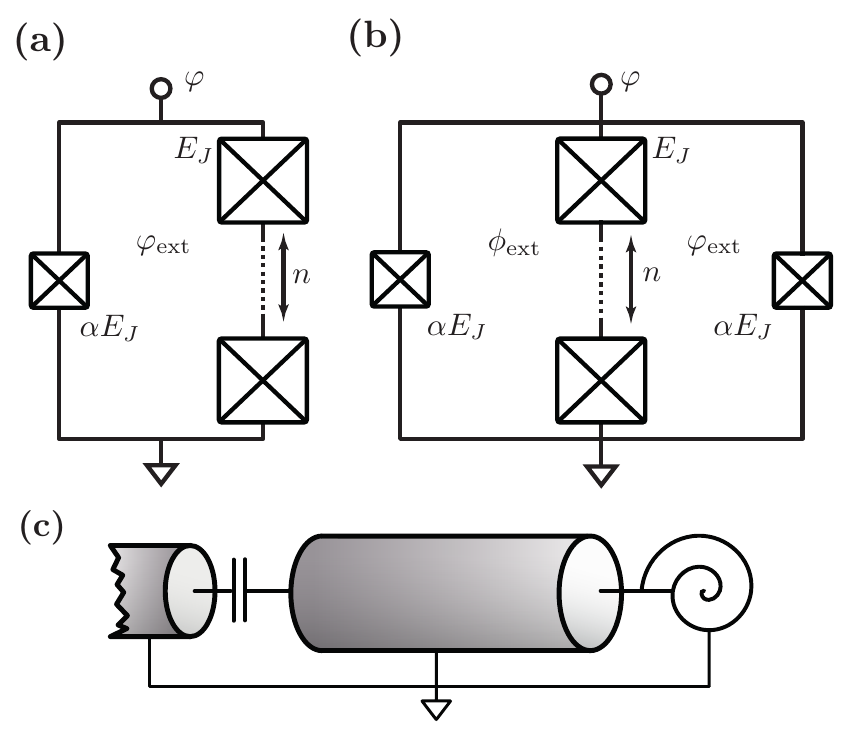}
    \caption{
    Circuit representation of the generalized SNAIL and the generalized ATS element after reduction to a single degree of freedom $\varphi$ in panels (a) and (b), respectively.
    (a) The inductive energy of the SNAIL can be tuned in situ by applying an external magnetic flux $\varphi_{\mathrm{ext}}$ through the loop formed by $n$ large Josephson junctions with energies $E_J$ and a single smaller junction with energy $\alpha E_J$ ($\alpha < 1$).
    (b) The ATS element can be understood as an inductively shunted SQUID or as a further generalization of the SNAIL with an additional loop formed by another small junction with energy $\alpha E_J$ \timo{(i.e., we assume both small junctions are identical)}. The added loop allows for further controllability after fabrication in comparison to the SNAIL.
    Both dipole elements permit further modification of the potential during fabrication through the parameters $n$ and $\alpha$.
    (c) Example for a single mode system: A transmission line resonator is terminated into a SNAIL at the right end and capacitively coupled to an input transmission line at the left end through which microwave signals for control can be fed (adapted from Ref.~\cite{hillmann_universal_2020}). 
    }
    \label{fig:snail-circuit}
\end{figure}
%
\section{Nonlinear photon interactions from asymmetric Josephson-Junction loops  \label{sec:motivation_hamiltonian}}
In microwave quantum optics, commonly referred to as circuit QED (cQED), nonlinear photon interactions are made possible by coupling the modes of interest to Josephson junctions acting as nonlinear, low-loss inductive elements with potential energy $U(\varphi) = E_J [1 - \cos \varphi]$, where $\varphi$ is the superconducting phase across the junction and $E_J$ is the Josephson energy~\cite{blais_circuit_2021}.
Arranging Josephson junctions in a loop configuration
allows for \emph{in situ} tuning of the potential through the external magnetic flux threading the loop as well as its parametric modulation~\cite{wustmann_parametric_2013, wustmann_nondegenerate_2017}.
The best known example is the dc SQUID in which a superconducting loop is interrupted by
two Josephson junctions with energies $E_{J, l}$ and $E_{J, r}$. 
In the symmetric case $E_{J, l} = E_{J, r}$, the parity of the potential remains even, and therefore any resulting nonlinear mixing process involves an even number of photons, e.g., four-wave mixing generated by a $\varphi^4$ term.
\fernando{A consequence of the even parity of the potential is the presence of (commonly) undesired nonlinear processes, such as self- and cross-Kerr interactions. These processes are always resonant (non-rotating) as they contain an equal number of creation and annihilation operators.
Thus, 
breaking the even symmetry of the Josephson potential 
allows, in principle, to engineer three-wave mixing processes (originating from a $\varphi^3$ term) with small or negligible residual Kerr terms.
}
These processes are of particular interest for the realization of quantum-limited Josephson parametric amplifiers~\cite{bergeal_phase-preserving_2010, frattini_3-wave_2017, frattini_optimizing_2018, sivak_kerr-free_2019}, but also for generating a tunable beam-splitter type interaction $g_{\mathrm{eff}}(t)(\hat{a}^{\dagger} \hat{b} + \hat{a} \hat{b}^{\dagger})$ between two modes $\hat a$ and $\hat b$.
\fernando{As demonstrated in Refs.~\cite{svensson_period_2018, chang_observation_2020},  \emph{time-dependent} odd parity contributions to the Josephson potential
can be obtained by flux-pumping an asymmetric dc SQUID ($E_{J, l} \neq E_{J, r}$). 
Nevertheless, the strength of the odd-parity terms is proportional to the small modulation amplitude. This results in four-mixing processes giving the leading nonlinear contribution to the potential over three-wave mixing ones.}\footnotetext[98]{To be precise, three-wave mixing has already been achieved with Josephson ring modulators (JRM)~\cite{bergeal_phase-preserving_2010, bergeal_analog_2010}.
However, the JRM has the drawback that it is a quadropole element and thus provides only a trilinear Hamiltonian $\varphi_X \varphi_Y \varphi_Z$ between three modes $X, Y$ and $Z$.}

To overcome these difficulties and achieve a pure $\varphi^3$ interaction, the Superconducting Nonlinear Asymmetric Inductive eLement (SNAIL)~\cite{frattini_3-wave_2017, frattini_optimizing_2018, sivak_kerr-free_2019, vool_engineering_2017, Note98} has been introduced.
The three-wave mixing capabilities of the SNAIL arise from an asymmetric arrangement of Josephson junctions in a loop configuration through which an external magnetic flux $\Phi_{\mathrm{ext}}$ is applied.
In particular, the loop is formed by $n$ large Josephson junctions in parallel with a single smaller junction with energies $E_J$ and $\alpha E_J$ $(\alpha < 1)$, respectively.
The SNAIL circuit is shown in \sfigref{fig:snail-circuit}{a}.
Neglecting the effects of capacitances across the junctions, the potential energy of the circuit in \sfigref{fig:snail-circuit}{a} can be reduced to a single degree of freedom $\varphi$ and becomes~\cite{frattini_3-wave_2017}
\begin{align}
    \label{eq:snail-potential}
    U_{\mathrm{SNAIL}}(\varphi)=-\alpha E_{J} \cos (\varphi)-n E_{J} \cos \left(\frac{\varphi_{\mathrm{ext}}-\varphi}{n}\right).
\end{align}
Here $\varphi$ denotes the superconducting phase across the small junction [see \sfigref{fig:snail-circuit}{a}], $\varphi_{\mathrm{ext}}=2 \pi \Phi_{\mathrm{ext}} / \Phi_{0}$
is the reduced applied magnetic  flux  and  $\Phi_0 = h / 2e$ is  the  (superconducting)  flux quantum.

The SNAIL dipole element can be further generalized by adding an additional single smaller junction parallel to the array of the $n$ large junctions, see \sfigref{fig:snail-circuit}{b}, which forms a second loop through which another reduced external magnetic flux $\phi_{\mathrm{ext}}$ can be applied.
The resulting device is called the Asymmetrically Threaded SQUID (ATS)~\cite{lescanne_exponential_2020}, as it can be understood as an inductively shunted symmetric SQUID.
If the above mentioned constraints are fulfilled, the potential energy of the ATS dipole element can be written as
\begin{align}
    U_{\mathrm{ATS}}(\varphi) = -2 \alpha E_{J} \cos(\varphi_{\Sigma}) \cos(\varphi + \varphi_{\Delta}) - n E_{J} \cos(\frac{\varphi}{n}),
\end{align}
where we have introduced $2 \varphi_{\Sigma} = \varphi_{\mathrm{ext}} + \phi_{\mathrm{ext}}$ and $2 \varphi_{\Delta} = \varphi_{\mathrm{ext}} - \phi_{\mathrm{ext}}$ which correspond to the sum and difference of the external fluxes applied through the loops, respectively.
In Ref.~\cite{lescanne_exponential_2020} the authors chose $n=5$ such that the Josephson array can be approximated by a standard inductive element and that the interaction is solely mediated through the term $ -2 \alpha  E_{J} \cos(\varphi_{\Sigma}) \cos(\varphi + \varphi_{\Delta})$.
The additional controllability obtained through the second loop allows to tune nonlinear couplings of the device \emph{in situ} while leaving the quadratic coupling, and thus effectively the frequency, untouched~\cite{miano_full_2021}.

The SNAIL and ATS have already been used in proposals~\cite{hillmann_universal_2020, grimsmo_quantum_2021} and experimental implementations~\cite{lescanne_exponential_2020,noguchi_fast_2020, noguchi_single-photon_2020} to realize linear and nonlinear interactions.
\timo{In the following we contribute further to the understanding of the nonlinear properties of such devices, especially to the renormalization of self-Kerr couplings through the cubic $\varphi^3$ interaction, by considering the embedding of these elements in single mode systems [\sfigref{fig:snail-circuit}{c}].}

\section{Deriving effective hamiltonians\label{sec:deriving_eff_hamiltonians}}
Today various methods to derive effective Hamiltonians exist~\cite{schrodinger_quantisierung_1926, james_effective_2007, bravyi_schriefferwolff_2011,paulisch_beyond_2014, zeuch_exact_2020}.
Especially in the fields of quantum optics or circuit quantum electrodynamics, the rotating wave approximation (RWA)~\cite{schleich_quantum_2001} is a commonly used tool to simplify the (analytical) analysis of quantum systems.
Formally, one argues that in the interaction picture the terms that are rapidly oscillating average out over relevant time-scales of the system and therefore do not contribute significantly to the dynamics.

\subsection{Time-averaged Hamiltonian}
The James' effective Hamiltonian method~\cite{james_quantum_2000, james_effective_2007, gamel_time-averaged_2010} is a particular efficient way to obtain leading order corrections to the RWA.
To this end, consider a system that is described in the rotating frame by a Hamiltonian $\hat{H}_{I}(t)$ that can be written as a sum of $N$ harmonic terms with frequencies $\omega_n > 0$, i.e., 
\begin{align}
    \label{eq:H_I_harmonic}
    \hat{H}_{I}(t)=\hat{h}_0 + \sum_{n=1}^{N} \hat{h}_{n}  \mathrm{e}^{-i \omega_{n} t} +\hat{h}_{n}^{\dagger} \mathrm{e}^{i \omega_{n} t},
\end{align}
and a time-independent, diagonal term $\hat{h}_0$.
Without loss of generality we can assume \timo{that the frequencies $\omega_n$ and $\omega_m$ of different harmonic terms satisfy} $\omega_n \neq \omega_m$, otherwise we can combine these terms into a single one.

As derived elsewhere~\cite{james_quantum_2000, james_effective_2007, gamel_time-averaged_2010}, the effective Hamiltonian that contains the leading order corrections to the RWA is given by
\begin{align}
    \label{eq:time_averaged_eff_H}
    \hat{H}_{\mathrm{eff}} = \tavg{\hat{H}_{I}(t)} - \frac{i}{2} \tavg{\comm{ \hat{H}_{I}^{'}(t)}{ \int_0^t \hat{H}_{I}^{'}(t') \dd{t'}}},
\end{align}
where the overline denotes the time-average, that is,
\begin{align}
    \tavg{\hat{H}(t)} = \lim_{T \to \infty} \frac{1}{T} \int_0^{T} \hat{H}(t) \dd{t},
\end{align}
and $\hat{H}^{'}_{I}(t) = \hat{H}_{I}(t) - \tavg{\hat{H}_{I}(t)}$ contains only the time-dependent part of $\hat{H}_I(t)$.
Inserting Eq.~\eqref{eq:H_I_harmonic} into Eq.~\eqref{eq:time_averaged_eff_H}, one finds
\begin{align}
    \label{eq:H_eff_short}
    \hat{H}_{\mathrm{eff}} = \hat{h}_0 + \sum_{n=1}^{N} \frac{1}{\omega_n} \comm{\hat{h}_n^{\dagger}}{\hat{h}_n},
\end{align}
as all other terms have a finite oscillation frequency and are averaged out.
Whether neglecting the oscillating terms that are of the form $\comm*{\hat{h}_m^{(\dagger)}}{\hat{h}_n}$ is valid, can be verified from the criteria for the rotating wave approximation with effective frequencies $\lvert \omega_m \pm \omega_n \rvert$.

\subsection{Full-diagonalizing Schrieffer-Wolff transformation \label{ssec:fd_schrieffer_wolff}}
The full-diagonalizing Schrieffer-Wolff (SW) transformation can be utilized to obtain 
corrections in arbitrary order to the RWA.
\timo{It can be applied to general Hamiltonians of the form}
\begin{align}
    \label{eq:h_free_plus_perturb}
    \hat{H} = \hat{H}_0 + \lambda \hat{V},
\end{align}

\timo{
where $\hat{H}_0$ is a free, diagonal Hamiltonian and $\hat{V}$ is a small (nonlinear) perturbation, i.e., the operator norm $\norm{V} = \max_{\ket{x}} \norm{V \ket{x}}$ should be smaller than half of the spectral gap $\Delta_0 / 2$ of $\hat{H_{0}}$~\cite{bravyi_schriefferwolff_2011}.
The problem of formally defining the term \emph{small perturbation} for bosonic systems where $\hat{V}$ is typically unbounded can be overcome when restricting the problem to a specific state manifold, e.g., states that have support only on the lowest $n$ Fock states.
Whether the higher-order corrections to the RWA obtained from the SW transformation accurately describe the system will therefore depend on the physically relevant states and is best determined self-consistently~\cite{kessler_generalized_2012}.
The role of the parameter $\lambda$ is solely to simplify order counting that indicates to which order in $\hat{V}$ the Hamiltonian~\eqref{eq:h_free_plus_perturb} is diagonal with respect to the basis of $\hat{H}_0$~\cite{Note44}.
\footnotetext[44]{If $\hat{V}$ contains multiple energy scales, it can be useful to introduce multiple parameters and $\lambda_i$ and split $\hat{V}$ into multiple terms, i.e., $\lambda \hat{V} \doteq \sum_i \lambda_i \hat{V}_i$.}
}



Corrections to the RWA are obtainted by perturbatively diagonalizing $\hat{H}$ order-by-order through a unitary transformation $\hat{H}_{\mathrm{eff}} = e^{\hat{S}} \hat{H} e^{-\hat{S}}$ generated by the anti-Hermitian operator $\hat{S}$.
To this end, one assumes a power series expansion of $\hat{S}$, that is,
\begin{align}
    \label{eq:SW_trafo_generator}
    \hat{S} = \sum_{n=1}^{\infty} \lambda^n \hat{S}^{(n)}.
\end{align}
As a result, the effective Hamiltonian can be written as~\cite{poletto_entanglement_2012}
\begin{align}
    \label{eq:schrieffer_wolff_series_expansion}
    \hat{H}_{\mathrm{eff}} &= e^{\hat{S}} \hat{H} e^{-\hat{S}} = \sum_{j=0}^{\infty} \frac{1}{j!} \left[ \sum_{n=1}^{\infty} \lambda^n \hat{S}^{(n)}, \hat{H}_{0} + \lambda \hat{V} \right]_j \\
    &= \sum_{m=0}^{\infty} \lambda^m \hat{H}^{(m)},
\end{align}
where $\comm*{\hat{X}}{\hat{Y}}_j = \comm*{\hat{X}}{\comm*{\hat{X}}{\hat{Y}}_{j-1}}$ denotes the iterated commutator with $\comm*{\hat{X}}{\hat{Y}}_0 = \hat{Y}$.
In general, the $m$-th order term $\hat{H}^{(m)}$ of the effective Hamiltonian takes the form
\begin{align}
    \label{eq:SW_Hm_term}
    \hat{H}^{(m)} &= \comm{\hat{S}^{(m)}}{\hat{H}_0} + \hat{V}^{(m)}\left(\lbrace \hat{S}^{(n)} \rbrace_{n=1}^{m-1}, \hat{H_0}, \hat{V}\right),
\end{align}
where the generalized potential operator $\hat{V}^{(m)}$ depends upon all terms of the generator $\hat{S}$ up to order $n = m - 1$ and we define $\hat{V}^{(1)} = \hat{V}$.
To lighten the notation, we make the dependencies of $\hat{V}^{(m)}$ implicit in the following.

Typically, one is interested in choosing $\lbrace \hat{S}^{(n)} \rbrace_{n=1}^{m-1}$ such that $\hat{H}_{\mathrm{eff}}$ is diagonal up to order $m$, that is, for all $n \leq m-1$ it must hold that,
\begin{align}
    \label{eq:SW_diag_condition}
    \comm{\hat{S}^{(n)}}{\hat{H}_0} + \hat{V}^{(n)}_N
    = 0,
\end{align}
and the subscript $N$ denotes the part of $\hat{V}^{(n)}$ that is off-diagonal.
We devised an iterative protocol to construct $\hat{S}$ at the operator level through the relation between the James' effective Hamiltonian method and second-order Schrieffer-Wolff perturbation theory.
This follows from the observation that the 
anti-derivative of the
interaction picture Hamiltonian $\hat{H}_I(t) = {\rm e}^{i \hat H_0 t} \hat V {\rm e}^{-i \hat H_0 t}$,
Eq.~\eqref{eq:H_I_harmonic}, evaluated at $t=0$ agrees with the first term $\hat{S}^{(1)}$ in the power series expansion~\eqref{eq:SW_trafo_generator}.
From this and the general form~\eqref{eq:SW_Hm_term} we show in \appref{appsec:generator_proof} that 
\begin{align}
    \label{eq:SW_Sm_ansatz}
    \hat{S}^{(m)} =\lim_{t' \to 0} i \int e^{i \hat{H}_0 t'} \hat{V}^{(m)} e^{- i \hat{H}_0 t'} \dd{t'},
\end{align}
formally diagonalizes $\hat{H}^{(m)}$ if the full Hamiltonian $\hat{H} = \hat{H}_0 + \hat{V}$ is time-independent.
Here, we denote with $\int f(x) \dd{x}$ the anti-derivative (or indefinite integral) of $f(x)$ and the integration constant is chosen to be zero.

Consider now the case that $\hat{H}_0$ is of the form
\begin{align}
    \label{eq-app:oscillator-hamiltonian}
    \hat{H}_{0} = \sum_j \omega_j \hat{a}^{\dagger}_j \hat{a}_j,
\end{align}
possibly composed of multiple bosonic modes $\comm*{\hat{a}_i}{\hat{a}^{\dagger}_j} = \delta_{ij}$ with bare resonance frequencies $\omega_j$
and that $\hat{V}$ is a polynomial function of bosonic annihilation and creation operators $\lbrace \hat{a}_{i}, \hat{a}^{\dagger}_i \rbrace$, that is, the Taylor expansion for $\hat{V}$ converges.
In this case the ansatz~\eqref{eq:SW_Sm_ansatz} provides an efficient approach to iteratively perform the full-diagonalization transformation order-by-order on the operator level, because $e^{i \hat{H}_0 t'} \hat{V}^{(m)}_N e^{- i \hat{H}_0 t'}$ can be expressed as a finite sum of harmonic terms [cf. Eq.~\eqref{eq:H_I_harmonic}].
We emphasize that this is in stark contrast to other formal solutions of Eq.~\eqref{eq:SW_diag_condition} which specify $\hat{S}^{(n)}$ only through its matrix elements~\cite{winkler_spin_2003, magesan_effective_2020}.
This distinction becomes increasingly relevant if the Hilbert space of the problem is large, for example, in continuous-variable systems, in order to avoid the time-consuming calculation of (all relevant) matrix elements.

\timo{
Finally, we note that expression~\eqref{eq:SW_Sm_ansatz} could also be obtained by transforming the Hamiltonian $\hat{H}$ [Eq.~\eqref{eq:h_free_plus_perturb}] into the rotating frame with respect to the free Hamiltonian $\hat{H}_0$ and applying time-dependent Schrieffer-Wolff perturbation theory~\cite{bukov_universal_2015} with appropriate boundary conditions.
Recently, connections between multiple perturbative methods based on the Lie commutator were formalized in Ref.~\cite{venkatraman_static_2021}.
}


\section{Analytical results\label{sec:analytical_results}}
\subsection{Single mode system \label{ssec:ana_self_kerr_renorm}}
Upon quantization, the Hamiltonian of a single mode resonator terminated in a SNAIL or an ATS [cf. \sfigref{fig:snail-circuit}{c}] is given by~\cite{hillmann_universal_2020}
\begin{align}
    \label{eq:SNAIL-H}
    {\color{blue}
    \hat{H} = \omega_r \hat{a}^{\dagger} \hat{a} + \sum_{n \geq 3} g_n \left( \hat{a} + \hat{a}^{\dagger} \right)^n
    }
\end{align}
\fernando{with $\omega_r$ the resonance frequency and 
$g_n$ the nonlinear couplings that will depend on the macroscopic parameters of the circuit.
In particular the coupling strengths are related to the dimensionless zero-point fluctuations (zpf) of the superconducting phase $\zpf$, i.e., $g_n \sim \zpf^n$. 
We perform perturbation theory with respect to $\zpf$. 
Our goal in this section is to find the leading-order correction to the self-Kerr nonlinearity $\propto g_4$.
Consequently, one can neglect any corrections arising from the quartic term $\propto g_4$ in leading order perturbation theory as the leading order correction is solely determined by the cubic anharmonicity $\propto g_3$.
This is equivalent to considering the  limit $g_4 \rightarrow 0$ for determining the effective Hamiltonian.
We will reintroduce these terms only at the end of the derivation in order to determine the self-Kerr free point.
}

Now, to apply the methods of \secref{sec:deriving_eff_hamiltonians} we first transform the Hamiltonian~\eqref{eq:SNAIL-H} into the interaction picture to obtain
\begin{align}
    \label{eq:H-SNAIL-int}
    \hat{H}_{I}(t) = g_3 \left( \hat{a} e^{-i \omega_r t} + \hat{a}^{\dagger}e^{i \omega_r t} \right)^3,
\end{align}
from which one can identify $\hat{h}_1 = 3 \hat{a}^{\dagger} \hat{a} \hat{a}^{\dagger}$ and $\hat{h}_2 = \hat{a}^{\dagger 3}$ with $\omega_1 = \omega_r$ and $\omega_2 = 3 \omega_r$, respectively.
Inserting those into Eq.~\eqref{eq:H_eff_short} we obtain the effective Hamiltonian in the absence of the quartic interaction which is given by
\begin{align}
    \hat{H}_{\mathrm{eff}}^{'} = - 30 \frac{g_3^2}{\omega_r}\left[ \hat{a}^{\dagger 2} \hat{a}^{2} + 2 \hat{a}^{\dagger} \hat{a} \right].
\end{align}
For small but finite $g_4$ we can reintroduce the diagonal terms originating from the quartic $\varphi^4$ interaction to obtain
\begin{align}
    \label{eq:SW-H-eff}
    \hat{H}_{\mathrm{eff}} \approx \delta \hat{a}^{\dagger} \hat{a}  + \frac{K}{2} \hat{a}^{\dagger 2} \hat{a}^2,
\end{align}
with
\begin{align}
    \delta &= 12 \left( g_4 - 5 \frac{g_3^2}{\omega_r} \right) \label{eq:freq-shift-one-mode}, \\
    K &= 12 \left( g_4 - 5 \frac{g_3^2}{\omega_r} \right) \label{eq:self-Kerr-one-mode},
\end{align}
which is accurate up to $\order*{\zpf^6}$, because there cannot be a diagonal contribution $\sim g_3 g_4 / \omega_r \sim \zpf^5$ \fernando{(as $\omega_r \sim \zpf^2$)}.
The reason is that the commutator of a cubic and quartic term will be off-diagonal and thus cannot directly renormalize the Kerr coupling.
\timo{As a result, we find that up to order $\order*{\zpf^6}$} the effective Hamiltonian $\hat{H}_{\mathrm{eff}}$ is given by the linear Hamiltonian $\hat{H}_0$ at the Kerr free point 
\begin{align}
    \label{eq:kerr-free-point}
    g_4^{\star} = 5 g_3^2 / \omega_r,
\end{align}
where the linear frequency shift $\delta$ [Eq.~\eqref{eq:freq-shift-one-mode}] vanishes as well.

\fernando{
The influence of higher-order corrections can be studied using the Schrieffer-Wolff transformation method.
Then the perturbative order we wish to consider determines how many non-linear
terms in the Hamiltonian~\eqref{eq:SNAIL-H} need to be taken into account.
While in conventional applications~\cite{frattini_optimizing_2018,lescanne_exponential_2020, noguchi_fast_2020}, 
it suffices to describe a SNAIL/ATS by 
the truncated Hamiltonian 
\begin{align}
    \label{eq:SNAIL-trunc}
    \hat{H} = \omega_r \hat{a}^{\dagger} \hat{a} + g_3 \left( \hat{a} + \hat{a}^{\dagger} \right)^3 + g_4 \left( \hat{a} + \hat{a}^{\dagger} \right)^4, 
\end{align}
which results from neglecting contributions of order $\mathcal{O}(\zpf^5)$ in~\eqref{eq:SNAIL-H},
this is not the case if one seeks an accurate perturbative expansion in the zero-point fluctuations $\zpf$.
However, the truncated Hamiltonian Eq.~\eqref{eq:SNAIL-trunc} presents a pedagogical toy model for which we have derived the corrections to $\hat H_{\rm eff}$~\eqref{eq:SW-H-eff} up to $\zpf^{8}$ using the method based on the Schrieffer-Wolff transformation introduced above.
Since for the next-to-leading order corrections one has to consider terms that are proportional to higher order powers of $\zpf$, one should no longer neglect higher-order nonlinearities in the initial Hamiltonian~\eqref{eq:SNAIL-H}, in this case, $g_5 (\hat{a} + \hat{a}^{\dagger})^5$ and $g_6 (\hat{a} + \hat{a}^{\dagger})^6$. 
Nevertheless, these results highlight  that for the goal of designing an effectively linear system it is not sufficient to only cancel the Kerr coupling, but also generalized higher-order Kerr terms $\sim \hat{a}^{\dagger \ell} \hat{a}^{\ell}, \ell \geq 3$.}

\begin{figure}
    \centering
    \includegraphics{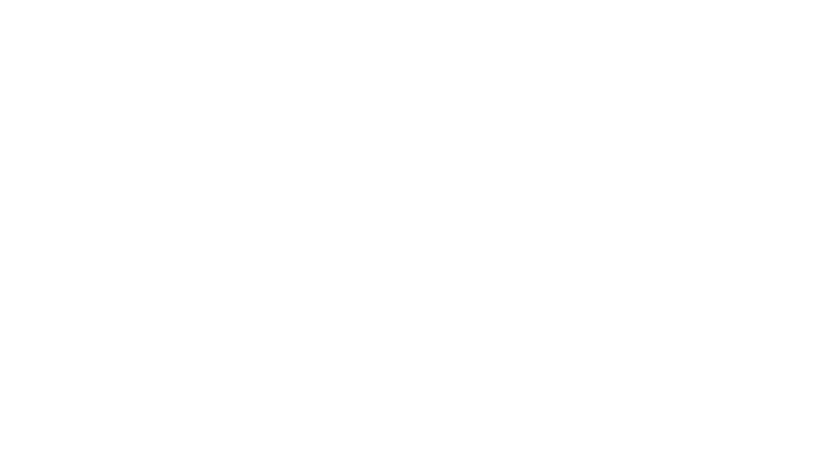}
    \caption{Diagrammatic visualization of processes that contribute to the effective Hamiltonian. (a) All processes that are effectively generated from a single-mode Hamiltonian that contains cubic and quartic nonlinearities, e.g. Eq.~\eqref{eq:SNAIL-H}, can be represented in terms of elementary vertices with three and four legs. Vertices are connected at their legs by red lines that represent Wick contractions that take place during normal ordering.
    (b) Some example diagrams with $2 \ell = 4$ uncontracted legs that contribute to the effective Kerr nonlinearity $K$. The first line includes the bare quartic vertex as well as diagrams corresponding to leading-order correction to the RWA (second-order perturbation theory). The second line shows diagrams that occur only in third-order as they contain three vertices each.}
    \label{fig:diagrammatic_rep}
\end{figure}

Lastly, in order to obtain a better understanding for the physical processes that contribute in $m$-th order to the effective Hamiltonian, we translated Eq.~\eqref{eq:SW_Hm_term} into a  qualitative diagrammatic picture, see \figref{fig:diagrammatic_rep}. Further details can be found in
\appref{app:diagrammatics}.
While this picture shares some common features with other diagrammatic techniques encountered in condensed matter physics~\cite{bravyi_schriefferwolff_2011, zinn-justin_quantum_2002}, we have so far not attempted to explicitly relate them.
We visualize terms that are cubic and quartic in the bosonic operators by bare interaction vertices with three and four legs, respectively.
Individual vertices are connected by red lines that represent Wick contractions.
Because every term in Eq.~\eqref{eq:SW_Hm_term} (for $m \geq 2$) involves at least a single commutator, only connected diagrams contribute. 
At order $m$, which equals the number of interaction vertices, each term is proportional to $g_3^{n_3} g_4^{n_4}/\omega^{m-1}_r$ with $n_3 + n_4 = m$ and contains at least $m-1$ Wick contractions.
Additional Wick contractions occur during the normal ordering procedure of the resulting boson string.
Through the normal ordering, additional nonlinear diagonal
contributions 
$\hat{a}^{\dagger \ell} \hat{a}^{\ell}$
are generated at any order $m$.
Here, $2 \ell$ denotes the number of uncontracted legs in a single diagram so that, for example, diagrams with four uncontracted legs ($\ell = 2$) correspond to a Kerr term $\propto \hat{a}^{\dagger 2} \hat{a}^{2}$, see also \sfigref{fig:diagrammatic_rep}{b}.
Note that all contributions contain an equal number of creation and annihilation operators because we do not consider any external drives.
Following this diagrammatic representation, it is straightforward to see why Eq.~\eqref{eq:SW-H-eff} cannot contain any contribution $\sim g_3 g_4 / \omega_r$. This is due to the fact that
the contraction of a cubic and and a quartic vertex always results in an odd number of uncontracted legs.
Formalizing this intuitive picture and deriving combinatorial rules for the numerical prefactors of each diagram is left for future work.

\section{Numerical analysis\label{sec:numerical_results}}
\subsection{Single mode system \label{ssec:num_self_kerr_renorm}}
In this section we will consider numerically the renormalization of the self-Kerr interaction to verify our analytic considerations above.
\fernando{
For this we are going to restrict to the truncated Hamiltonian~\eqref{eq:SNAIL-trunc} as a toy model
of a SNAIL/ATS assuming that higher-order nonlinearities are identical to zero, i.e., $g_n =0$, for $n \geq 5$.
While in principle, the $g_n$ couplings in~\eqref{eq:SNAIL-H} can be derived from a microscopic model of the circuit,
their dependence on the circuit parameters and the external flux is non-trivial and one would also need to judiciously choose the operation point of the device for a particular application. This is out of the scope of this work. 
}

The numerical methods will also yield an effective way to study the influence of higher than leading order corrections.
To study whether our considerations above are meaningful we will consider different approaches.

\subsubsection{Hamiltonian diagonalization\label{sssec:hamiltonian_diagonalization}}


A straightforward approach to analyze the effective Kerr nonlinearity of the system is by numerical diagonalization to obtain the (numerically) exact eigenenergies $E_n$.
We are interested in the functional dependence of $E_n$ on the quantum number $n$.
For a harmonic oscillator it is well known that $E_n = n \omega_r + 1/2$ is a linear function of $n$ while for the Kerr oscillator we expect $E_n$ to be a quadratic polynomial in $n$.
By considering the difference $\Delta E_n = E_n - n E_1, n \geq 1$,  we expect to remove the (dominating) linear dependence and ideally observe a flat curve at the Kerr-free point defined by Eq.~\eqref{eq:kerr-free-point}.

We choose the parameters $\omega_r \dtwopi = \SI{6}{\GHz}$ and $g_4 \dtwopi = \SI{2}{\kHz}$ for the simulation based on an experiment which achieves the controlled release of multiphoton quantum states~\cite{pfaff_controlled_2017, axline_-demand_2018, burkhart_error-detected_2020}.
Here, the full fourth-order nonlinearity is a necessary ingredient to engineer a frequency conversion interaction. Nevertheless, the residual nonlinear terms that arise from the latter potential limit the approach to states of the form $\ket{\psi} = \sum_{n=0}^{5} c_n \ket{n}$. 
The reason is that the frequency of higher Fock states is shifted beyond the frequency conversion bandwidth.

\begin{figure}[!ht]
	\centering
    \includegraphics{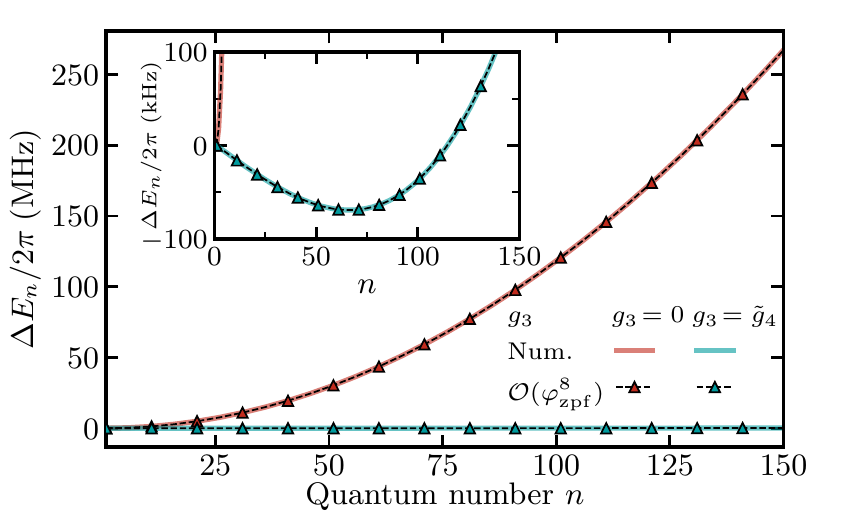}
	\caption{Energy level difference $\Delta E_n = E_n - n E_{1}$ for the Hamiltonian~\eqref{eq:SNAIL-H} plotted against the quantum number $n \geq 1$.
	Main panel: The blue curve for $g_3 = \tilde{g}_4 =\sqrt{g_4 \omega_r / 5}$ appears flat while the red curve for $g_3 = 0$ has an approximately quadratic dependence on $n$.
	Solid lines represent the numerically (Num.) obtained eigenvalues and
	black dashed lines with markers are the analytically obtained eigenenergies from $m$-th-order perturbation theory which is accurate up to order \timo{$\order{\zpf^{8}}$}, see \appref{app:schrieffer_wolff}.
	Inset: The zoom into $\lvert \Delta E_n \rvert \leq \SI{100}{\kHz}$ shows that $\Delta E_n$ for $g_3 = \sqrt{g_4 \omega_r / 5}$ is not flat. 
	Note the different units of the y-axis between the main panel and the inset.
	Other parameters are $\omega_r \dtwopi = \SI{6}{\GHz}$ and $g_4 \dtwopi  = \SI{2}{\kHz}$.}
	\label{fig:spectrum}
\end{figure}

In \figref{fig:spectrum} we show the difference $\Delta E_n$ for $n \geq 1$ for $g_3 = 0$ (red solid line) and $g_3 = \sqrt{g_4 \omega_r / 5}$ (blue solid line).
We observe that for $g_3 = \sqrt{g_4 \omega_r / 5}$ the curve is practically flat with respect to the curve for $g_3 = 0$ that shows the expected quadratic increase.
A zoom into the region where $\lvert \Delta E_n \rvert \leq \SI{100}{\kHz}$ (see inset) reveals that for the optimal choice of $g_3$ the curve is not completely flat.
Instead, $\Delta E_n$ is decreasing below $0$ until $n \approx 65$ after which it starts to increase again and eventually crosses the zero line again.
Using higher-order perturbation theory, as described in \appref{app:schrieffer_wolff}, we can verify that this behavior is mostly due to generalized Kerr terms $\hat{a}^{\dagger 3} \hat{a}^3$ that are not canceled in our approach.
\timo{The eigenenergies obtained from this approach are shown as thin black dashed lines with colored triangular markers.
We observe excellent agreement for the case of $g_4 \dtwopi = \SI{2}{\kHz}$ between our analytical results and the numerical exact diagonalization.
The agreement for larger quantum numbers $n$ at this order in perturbation theory is going to reduce if higher-order processes become more relevant, that is, when the ratio $g_4 / \omega_r$ increases.}
Note that for higher (drive-activated) photon numbers such as typically observed in SNAIL-based parametric amplifiers~\cite{frattini_3-wave_2017}, other perturbative methods are more appropriate~\cite{petrescu_accurate_2021}.

It is noteworthy that in principle the choice of $g_3$ can be optimized in order to minimize nonlinear effects up to some photon number $n_{\mathrm{max}}$.
Because in many experimental applications today only a fraction of the infinite dimensional oscillator Hilbert space is explored, e.g., only the first $n_{\mathrm{max}}$ Fock states~\cite{wang_efficient_2020}, one could minimize the nonlinearity up to this cut-off $n_{\mathrm{max}}$ for practical applications.
We explore this possibility further in \secref{sssec:kerr_oscillations}.

\fernando{
Effects of these higher-order Kerr terms cannot be canceled within our simplified $\varphi^3 + \varphi^4$ model.
To achieve this, one would require an additional free parameter, for example, coming from a $\varphi^5$ interaction.
Indeed, our model neglects these terms from the beginning as they are typically insignificant if the relevant quantum states have support only in the low excitation sector of the full Hilbert space.
}

\subsubsection{Kerr oscillations \label{sssec:kerr_oscillations}}
To analyze to which extent the resulting nonlinearity observed in \figref{fig:spectrum} 
is relevant 
and whether it is possible to optimize the value of $g_3$ further, we consider the evolution of a coherent state $\ket{\alpha_0}$ under the Hamiltonian~\eqref{eq:SNAIL-H}.
This is motivated by the observation that the expectation value of the field amplitude $ \lvert \langle \hat{a} \rangle_{\ket{\alpha_0}} \rvert (t)$ 
of a coherent state $\ket{\alpha_0}$ evolving under the Kerr oscillator (KO) Hamiltonian
\begin{align}
    \label{eq:Kerr-oscillator}
    \hat{H}_{\mathrm{KO}} = \omega_r \hat{a}^{\dagger} \hat{a} + \frac{K}{2} \hat{a}^{\dagger} \hat{a}^{\dagger} \hat{a} \hat{a},
\end{align}
will return to its initial value $\lvert \alpha_0 \rvert$ after a time $T = 2 \pi / K = \pi / 6 g_4$~\cite{yurke_dynamic_1988, kirchmair_observation_2013, elliott_designing_2018}.
Hence, the period of these oscillations is a direct indicator of the effective Kerr nonlinearity $K$.
Based on our analysis above we expect that there exists an optimal value of $g_3$ for which we have $T \rightarrow \infty$, i.e., negligible Kerr nonlinearity, for given $g_4$ and $\omega_r$.

\begin{figure}
    \centering
    \includegraphics{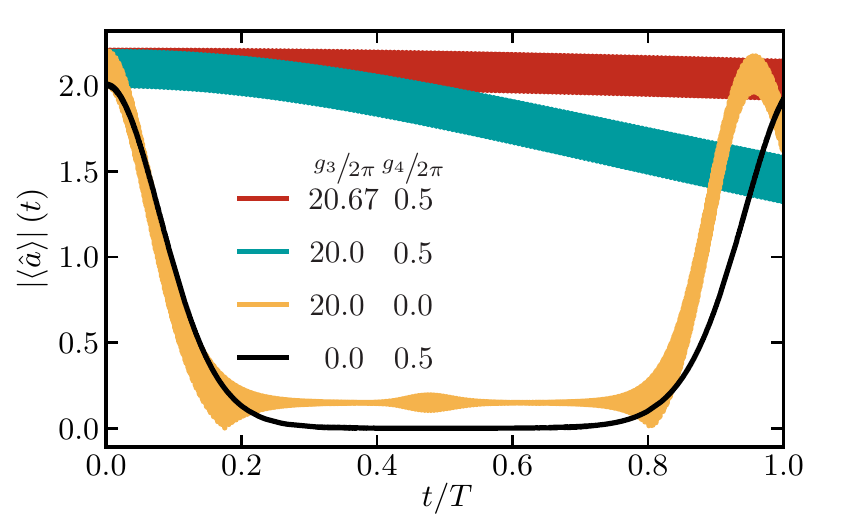}
    \caption{Cancellation of Kerr oscillations through the interplay of cubic and quartic interactions for a single mode system. The figure shows the unitary time evolution of the expectation value $ \lvert \langle \hat{a} \rangle_{\ket{\alpha_0}} \rvert$ 
    under the Hamiltonian~\eqref{eq:SNAIL-H} for different combinations of the nonlinear couplings $g_3$ and $g_4$. Other parameters are $\omega_r \dtwopi = \SI{4}{\GHz}$, $\alpha_0 = 2$
    and $T = \pi / 6 g_4^{\star}$ with $g_4^{\star} \dtwopi = \simhz{0.5}$  is fixed for all simulations.}
    \label{fig:kerr_oscillations}
\end{figure}

\figref{fig:kerr_oscillations} shows the time evolution of  $\vert \langle \hat a \rangle \vert$
for an initial coherent state $\ket{\alpha_0}$ with $\alpha_0 = 2$ in a resonator with frequency $\omega_r \dtwopi = \SI{4}{\GHz}$ for various combinations of couplings $g_3$ and $g_4$.
The case of a pure quartic interaction with $g_4 \dtwopi = \SI{0.5}{\MHz}$ and $g_3\dtwopi =\SI{0}{\MHz}$ shows the expected collapse and revival of the initial state within one period $T$.
A similar oscillation period can be obtained by choosing $g_3\dtwopi =\SI{20}{\MHz}$ and $g_4 \dtwopi = \SI{0}{\MHz}$ (orange line) which results in the same value for $K$ based on Eq.~\eqref{eq:self-Kerr-one-mode}.
The simulation with these parameters shows also high-frequency oscillations due to the fully off-resonant nature of the cubic interaction.
Based on Eq.~\eqref{eq:kerr-free-point} we expect vanishing Kerr oscillations ($T \to \infty$) for $g_3\dtwopi =\SI{20}{\MHz}$ and $g_4 \dtwopi = \SI{0.5}{\MHz}$ which is shown by the blue line in \figref{fig:kerr_oscillations}.
While we clearly observe an increase in the period of collapse and revival $T$, we also observe a reduction of the expectation value $\vert \langle \hat a \rangle \vert$.
As described in \appref{app:oscillation_average}, it is possible to optimize the value of $g_3$ to further reduce the decay of $\vert \langle \hat a \rangle \vert$.
Numerically we find that the parameter pair $g_3\dtwopi =\SI{20.67}{\MHz}$ and $g_4 \dtwopi = \SI{0.5}{\MHz}$ (red line) gives the best results.
Further intuition about this result can be gained from higher-order perturbation theory which results in an ``improved'' Kerr-free point that partially accounts for the discrepancy between the first-order result ($g_3\dtwopi =\SI{20}{\MHz}$) and the numerically optimized value ($g_3\dtwopi =\SI{20.67}{\MHz}$).
\fernando{At the improved point,
in order to reduce a positive contribution from $c_3 \hat{a}^{\dagger 3} \hat{a}^3$, it is actually beneficial
to have a small negative value of $K$ which accounts for the above discrepancy (see also \appref{app:schrieffer_wolff}).}

\subsubsection{Fidelity of cubic phase state preparation \label{sssec:cpg_fidelity}}
\fernando{Because a major motivation to utilize the interplay of
higher-order non-linearities}
is to improve gate operation in quantum information processing, we will consider here the cancellation of the effective Kerr nonlinearity during the operation of a cubic phase gate to prepare the cubic phase state~\cite{gottesman_encoding_2001}.
The cubic phase gate $\hat V(\gamma) = \exp(i \gamma (\hat{a} + \hat{a}^{\dagger})^3/ \sqrt{8} ) = \exp(i \gamma \hat{q}^3)$
together with a symplectic (Gaussian) gate can be used to implement the non-Clifford \qgate{T} on the GKP encoding~\cite{gottesman_encoding_2001}.
On the other hand, it is also possible to use the cubic phase state $\ket{\gamma, r} = e^{i \gamma \hat{q}^3} e^{(r/2)(\hat{a}^{\dagger 2} - \hat{a}^2)} \ket{0}$ as a resource state to implement $\hat{V}(\gamma)$ through gate teleportation.
To prepare a cubic phase state from an initial vacuum squeezed state $\ket{0, r} = e^{(r/2)(\hat{a}^{\dagger 2} - \hat{a}^2)} \ket{0}$ we consider a Hamiltonian of the form
\begin{align}
    \label{eq:H-CPG-kerr-cancel}
    \hat{H}(t) =&\, \omega_r \hat{a}^{\dagger} \hat{a} + \tilde{g}_3^{\mathrm{ac}}(t) \left(\hat{a} + \hat{a}^{\dagger} \right)^3 + g_3^{\mathrm{dc}} \left(\hat{a} + \hat{a}^{\dagger} \right)^3 \nonumber \\
    &+ g_4^{\mathrm{dc}} \left(\hat{a} + \hat{a}^{\dagger} \right)^4.
\end{align}
This results from the Hamiltonian Eq.~\eqref{eq:SNAIL-H} 
by considering a time-dependent $g_3$ coefficient $g_3(t) = g_3^{\rm dc} + \tilde{g}_3^{\mathrm{ac}}(t)$ in which the time-dependent part is chosen as $\tilde{g}_3^{\mathrm{ac}}(t) = g_3^{\mathrm{ac}} \left[\cos(\omega_r t) + \cos(3 \omega_r t) \right]$
in order to stabilize the full cubic interaction~\cite{hillmann_universal_2020}.
We consider evolution for a time $\tau = 2 \gamma / \sqrt{8} g_3^{\mathrm{ac}}$ to obtain a cubic phase state with cubicity $\gamma = 0.1$ which is sufficient for implementing the \qgate{T} gate~\cite{konno_non-clifford_2021}.
The effective Kerr nonlinearity, resulting from the interplay of the terms proportional to $g_3^{\mathrm{dc}}$ and $g_4^{\mathrm{dc}}$, will lead to a state preparation error
\begin{align}
\mathcal{E} = 1 - \lvert \braket{\gamma, r}{\psi(\tau)}\rvert^2, 
\end{align}
with $\ket{\psi(\tau)}$ the state after the unitary evolution of the initial state under the Hamiltonian~\eqref{eq:H-CPG-kerr-cancel} for a time $\tau$.
Based on Eq.~\eqref{eq:kerr-free-point} we expect that there is an optimal choice for $g_3^{\mathrm{dc}}$ that minimizes the error $\mathcal{E}$.

\begin{figure}[!ht]
    \centering
    \includegraphics{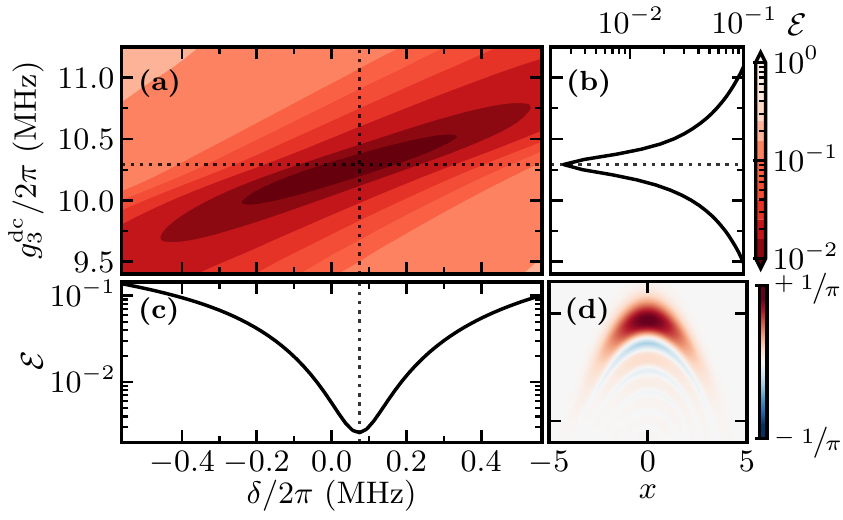}
    \caption{(a)~Colormap of the state preparation error $\mathcal{E}$ in the $(\delta, g_3^{\mathrm{dc}})$ parameter space obtained from the unitary evolution of an initial vacuum squeezed state $\hat{S}(-r)\ket{0}$ ($r=0.69$) under Hamiltonian~\eqref{eq:H-CPG-kerr-cancel} for a time $\tau$ such that $\gamma=0.1$. (b)~Cross-section of the gate error along the optimal choice of $\delta$ (vertical dotted line in (a)). (c)~Cross-section of the gate error along the optimal choice of $g_3^{\mathrm{dc}}$ (horizontal dotted line in \textbf(a)).
   (d)~Wigner distribution $W(x, p)$ at the optimal point in the $(\delta, g_3^{\mathrm{dc}})$  parameter space.
    Other parameters are: $\omega_r \dtwopi  = \SI{4}{\GHz}$, $g_4^{\mathrm{dc}} \dtwopi  = \SI{0.125}{\MHz}$ and $g_3^{\mathrm{ac}}  \dtwopi = \SI{0.25}{\MHz}$.}
    \label{fig:cpg-kerr-cancel}
\end{figure}

We test the cancellation of Kerr nonlinearity as well as the cancellation of the induced frequency shift [Eq.~\eqref{eq:freq-shift-one-mode}] by modifying $\omega_r \rightarrow \Tilde{\omega}_r = \omega_r + \delta$.
\sfigref{fig:cpg-kerr-cancel}{a} shows the gate error $\mathcal{E}$ as a colormap of the $(\delta \dtwopi , g_3^{\mathrm{dc}} \dtwopi )$ parameter space for $\omega_r \dtwopi  = \SI{4}{\GHz}$, $g_4^{\mathrm{dc}} \dtwopi  = \SI{0.125}{\MHz}$ and drive amplitude $g_3^{\mathrm{ac}} \dtwopi  = \simhz{0.25}$.
These parameters constitute a set of experimentally realistic parameters as discussed in Ref.~\cite{hillmann_universal_2020}.
\sfigref{fig:cpg-kerr-cancel}{b} and \sfigref{fig:cpg-kerr-cancel}{c} show cross-sections along the optimal choice of $\delta$ and $g_3^{\mathrm{dc}}$, respectively.
We find that their values are higher than our expectations based on Eq.~\eqref{eq:kerr-free-point}, i.e., $g_3^{\mathrm{dc}} = \SI{10}{\MHz}$ and $\delta = 0$, but in the vicinity of these predictions.
Comparing to the expected frequency shift of the resonator in the absence of the cubic nonlinearity $\delta \dtwopi  = 12 g_4^{\mathrm{dc}} \dtwopi  = \SI{1.5}{\MHz}$, we observe a significant suppression of $\delta$ as well.
The Wigner distribution $W(x, p)$ of the final state $\ket{\psi(\tau)}$ at the optimal point in the $(\delta, g_3^{\mathrm{dc}})$ parameter space is shown in \sfigref{fig:cpg-kerr-cancel}{d} and shows no sign of Kerr-induced distortion as expected from the gate error $\mathcal{E} < \num{e-2}$ at that point.



\section{Discussion and conclusion\label{sec:discussion_conclusion}}

In this work we have derived effective Hamiltonians for a class of
interacting models describing the quantum regime of superconducting circuits consisting of one or two loops interrupted by several non-identical Josephson junctions, 
namely the SNAIL and ATS circuits depicted in \sfigref{fig:snail-circuit}{a} and \sfigref{fig:snail-circuit}{b} respectively.
These configurations allow to break the even symmetry of the Josephson potential from where Hamiltonians that contain nonlinear odd and even order interactions can be obtained.
Using second-order perturbation theory, we obtain expressions for the frequency shift as well as the Kerr coupling of a single mode bosonic system.
\fernando{
We confirm our analytical expressions through numerical experiments 
in a simplified model of a SNAIL/ATS containing only cubic and quartic nonlinearities.
We show that effects of parasitic Kerr-type interactions can be strongly suppressed through the interplay of 
the above nonlinear photon-photon 
interactions.}
The required coupling strengths to achieve this cancellation are found to be in the vicinity of the values expected from the perturbative calculations.
Our results show that it is possible to increase the effective rates of linear and nonlinear operations without increasing the rates of parasitic nonlinear coherent interactions such as static self-Kerr couplings.
More generally, our work highlights the idea of harnessing off-resonant interactions that are ubiquitous in cQED systems as a mean to engineer quantum systems on the Hamiltonian level.
To do this efficiently, we provided as a computational tool an iterative algorithm that allows one to obtain the effective Hamiltonian analytically on the operator level up to arbitrary order in the off-resonant perturbation.
This algorithm builds on the formal solution for the generator of the Schrieffer-Wolff transformation derived here and is applicable beyond the models studied in this work.

Our work relates to studies that intend to engineer high-fidelity quantum operations for bosonic qubits~\cite{rosenblum_fault-tolerant_2018, wang_photon-number-dependent_2021, zhang_drive-induced_2021}.
While these works build upon particularly engineered external drives within the rotating wave approximation, our work consciously exploits nonlinear photon-photon interactions that are dynamically generated and appear due to the inclusion of off-resonant terms.
In particular, by providing an efficient method to compute these processes in arbitrary order, our work opens up the possibility to include off-resonant terms consistently, as it is now routinely done to engineer high-fidelity qubit operations~\cite{yan_tunable_2018, magesan_effective_2020, zeuch_exact_2020} in conventional architectures.


We note that our numerical simulations do not account for dissipative effects during the evolution of the quantum system.
However, simulating only unitary evolution is justified by the fact that the intention of the numerical simulations is to verify the validity of our analytical calculations which have been obtained on the level of the Hamiltonian.
The derivation of an effective master equation is left for future work.

Besides the above, numerous extensions of our work remain to be explored.
A first step can be to extend the results to time-dependent, multimode systems, such that it is possible to include (near-) resonant, drive-activated terms which will contribute to the renormalization of the nonlinear couplings depending on the drive strength~\cite{petrescu_lifetime_2020, petrescu_accurate_2021}.
\timo{
Ref.~\cite{petrescu_accurate_2021-1} has derived relevant expressions for this case already, but the interpretation of their results is limited to the parameter regime that is relevant for conventional (discrete-variable) quantum computing.
}
Furthermore, it is interesting to examine the question whether it is possible to suppress the effects of generalized Kerr-type interactions, perhaps through a single or multiple off-resonant drives~\cite{petrescu_accurate_2021, zhang_drive-induced_2021}.
It is possible to address this question from at least two angles.
On the one hand, it is possible to aim for the explicit cancellation of these nonlinear couplings order-by-order.
On the other hand, similar to our numerical optimization approach, through the effective cancellation of nonlinear effects for each Fock state, by engineering nonlinear couplings with alternating signs.
Lastly, even though (partial) cancellation of the Kerr nonlinerity has been observed in SNAIL parametric amplifiers~\cite{frattini_optimizing_2018, sivak_kerr-free_2019}, the experimental demonstrations in the context of a quantum information processing task remains.


\begin{acknowledgments}
T.H. acknowledges the financial support from the  Chalmers  Excellence  Initiative Nano.
F.Q. acknowledges the financial support from the Knut and Alice Wallenberg Foundation through the Wallenberg Centre  for  Quantum  Technology  (WACQT).
\end{acknowledgments}

\appendix

\section{Formal solution of the generator terms $\hat{S}^{(m)}$ \label{appsec:generator_proof}}
In this Appendix we show that the operator-valued equation
\begin{align}
    \label{eq-app:fd_schrieffer_wolff}
    \comm{\hat{S}^{(m)}}{\hat{H}_0} + \hat{V}^{(m)}_N\left(\lbrace \hat{S}^{(n)} \rbrace_{n=1}^{m-1}, \hat{H_0}, \hat{V}\right) = 0,
\end{align}
is formally solved by
\begin{align}
    \label{eq-app:SN-ansatz}
    \hat{S}^{(m)} = \lim_{t' \to 0} \left[ i \int e^{i \hat{H}_0 t'} \hat{V}^{(m)} e^{- i \hat{H}_0 t'} \dd{t'} \right].
\end{align}
Here, $\hat{V}^{(m)}$ is recursively defined through the series expansion given by equation~\eqref{eq:schrieffer_wolff_series_expansion} in \secref{ssec:fd_schrieffer_wolff} of the main text and we made the dependencies on lower-order terms implicit to lighten the notation.
Note that Eq.~\eqref{eq-app:SN-ansatz} contains the full generalized potential operator $\hat{V}^{(m)} = \hat{V}^{(m)}_S + \hat{V}^{(m)}_N $ while in Eq.\eqref{eq-app:fd_schrieffer_wolff} only the non-diagonal terms $\hat{V}^{(m)}_N$,  indicated by the subscript $N$, contribute.
However, the difference is inconsequential, because the diagonal (secular) part of $\hat{V}^{(m)}_S$ commutes with $\hat{H}_0$ as both operators are diagonal.

For the following it is useful to recall the relation of the commutator with a parameter derivative, that is, for a parameter $s$ one finds the relation
\begin{align}
    \label{eq-app:hadamard_formula}
    \dv{s} \left(e^{s \hat{A}} \hat{B} e^{-s \hat{A}}\right) = \left[\hat{A}, e^{s \hat{A}} B e^{-s \hat{A}} \right],
\end{align}
for two non-commuting operators $\hat{A}$ and $\hat{B}$ through the application of the product formula for derivatives.
Thus, by inserting the ansatz~\eqref{eq-app:SN-ansatz} into the first term of Eq.~\eqref{eq-app:fd_schrieffer_wolff} and using Eq.~\eqref{eq-app:hadamard_formula} with $s = i t'$, we have
\begin{align}
    \label{eq-app:proof_explicit_generator}
    \hspace{-10pt} \comm{\hat{S}^{(m)}}{\hat{H}_{0}} &= \lim_{t' \to 0} \comm{i \int e^{i \hat{H}_0 t'} \hat{V}^{(m)}_N e^{- i \hat{H}_0 t'} \dd{t'}}{\hat{H}_{0}} \\
    &= \dv{t'} \lim_{t' \to 0} \left[- \int e^{i \hat{H}_0 t'} \hat{V}^{(m)}_N e^{- i \hat{H}_0 t'} \dd{t'} \right] \\
    &= - \lim_{t' \to 0} e^{i \hat{H}_0 t'} \hat{V}^{(m)}_N e^{- i \hat{H}_0 t'} \\
    &= - \hat{V}^{(m)}_N,
\end{align}
as the limit commutes with the derivative and all operations are linear.
Thus, we have 
formally
shown that the ansatz~\eqref{eq-app:SN-ansatz} always 
solves the operator-valued equation~\eqref{eq-app:fd_schrieffer_wolff}
\footnote{After completing this work, we noticed that 
the authors of Ref.~\cite{sigmund_u-matrix_1974} found a different method to calculate $\hat S^{(1)}$.
Nevertheless, to the best of our knowledge their approach has not been  generalized to an arbitrary order in the perturbation. }.

We remark that its possible to relate Eq.~\eqref{eq-app:SN-ansatz} to previously known expressions for $\hat{S}^{(m)}$ on the level of its matrix elements~\cite{poletto_entanglement_2012}.
Projecting on the eigenstates $\ket{k}$ and $\ket{l}$ of the Hamiltonian $\hat{H}_0$ and performing the integral, we obtain
\begin{align}
    \label{eq-app:Sn_matrix_elements}
    \mel{k}{\hat{S}^{(m)}}{l} &= \lim_{t' \to 0} \left[ i \int e^{i (E_k - E_l) t'} \dd{t'} \mel{k}{\hat{V}^{(m)}_N}{l} \right] \\
    &= \frac{\mel{k}{\hat{V}^{(m)}_N}{l}}{E_k - E_l},
\end{align}
where we denoted the eigenenergies of the eigenstates $\ket{k}$ and $\ket{l}$ by $E_k$ and $E_l$, respectively.


\section{Analytical results for the single mode system \label{app:schrieffer_wolff}}
In this Appendix we present the analytical results for the full-diagonalizing Schrieffer-Wolff transformation for the single mode system discussed in \secref{ssec:ana_self_kerr_renorm} of the main text up to terms $\order{\zpf^6}$.
For convenience, we repeat the Hamiltonian~\eqref{eq:SNAIL-H} which is given by
\begin{align}
    \label{eq-app:SNAIL-H}
    \hat{H} = \omega_r \hat{a}^{\dagger} \hat{a} + g_3 \left( \hat{a} + \hat{a}^{\dagger} \right)^3 + g_4 \left( \hat{a} + \hat{a}^{\dagger} \right)^4.
\end{align}
While the leading order term $\hat{S}^{(1)}$ of the generator $\hat{S}$ is quickly obtained by hand, that is,
\begin{align}
    \hat{S}^{(1)} &= \frac{g_3}{\omega_r} \left( \frac{1}{3} \hat{a}^{\dagger 3} + 3 \hat{a}^{\dagger} \hat{a} \hat{a}^{\dagger} - \mathrm{H.c.} \right) \nonumber \\
    &+ \frac{g_4}{\omega_r} \left( \frac{1}{4} \hat{a}^{\dagger 4} + 2 \hat{a}^{\dagger 3} \hat{a} + 3 \hat{a}^{\dagger 2} - \mathrm{H.c.} \right), \label{eq-app:S1_generator}
\end{align}
the calculations of commutators become quickly tedious and it is useful to recast the iterative procedure into a computer algebra system.
The code that is written in Maple~\footnote{Maple 2021. Maplesoft, a division of Waterloo Maple Inc., Waterloo, Ontario.} and  is available for public use~\footnote{The code is available at \url{https://doi.org/10.5281/zenodo.5091424}.} and is straightforwardly generalized to arbitrary potential terms $\hat{V}$ or multiple modes.
We omit the intermediate results as well as the explicit expression for $\hat{S}^{(2)}$ and $\hat{S}^{(3)}$ as they are rather lengthy.
Instead, we specify the diagonal terms of the effective Hamiltonian
\begin{align}
    \hat{H}_{\mathrm{eff}}^{\mathrm{SW}} = \sum_{m=0}^4 \hat{H}^{(m)} \approx \sum_{n=0}^{\timo{3}} c_n \hat{a}^{\dagger n} \hat{a}^n,
\end{align}
in terms of the diagonal coefficients $c_n$ containing terms that are most proportional to $\zpf^6$.
They are given by
\begin{align}
    c_{3} &= -\frac{68 g_{4}^{2}}{\omega_r}  
    +\frac{1800 g_{3}^{2} g_{4}}{\omega_r^{2}}  
    +\frac{-2820 g_{3}^{4}}{\omega_r^{3}} \\ 
    c_{2} &= 6 g_{4} +\frac{-30 g_{3}^{2} - 306 g_{4}^{2}}{\omega_r}+\frac{8100 g_{3}^{2} g_{4}}{\omega_r^{2}}+\frac{-12690 g_{3}^{4}}{\omega_r^{3}}, \\
    c_{1} &= 12 g_{4}  +\frac{-60 g_{3}^{2} -288 g_{4}^{2}}{\omega_r}+\frac{6768 g_{3}^{2} g_{4} }{\omega_r^{2}}+ \frac{-10320 g_{3}^{4}}{\omega_r^{3}}, \\
    c_0 &= 3 g_{4} +\frac{-11 g_{3}^{2}  -42 g_{4}^{2}}{\omega_r}+\frac{684 g_{3}^{2} g_{4} }{\omega^{2}_r}+\frac{-930 g_{3}^{4}}{\omega^{3}_r}.
\end{align}
We used these coefficients in order to analytically compute the energy level difference $\Delta E_n = E_n - n E_{1}$ shown in \figref{fig:spectrum} of the main text.

\section{Details on the diagrammatic picture\label{app:diagrammatics}}
Here we give further details on the qualitative diagrammatic picture that was introduced in \figref{fig:diagrammatic_rep} of the main text.
To this end, we will explain some of the ``rules'' on a simplified example that contains only the cubic interaction, that is, we consider the Hamiltonian
\begin{align}
    \hat{H}_{3} = \underbrace{\omega_r \hat{a}^{\dagger} \hat{a}}_{= \hat{H}_0} + \underbrace{g_3 (\hat{a}^{\dagger 3} + 3 \hat{a}^{\dagger} \hat{a} \hat{a}^{\dagger} + 3 \hat{a} \hat{a}^{\dagger} \hat{a} + \hat{a}^{3})}_{= \hat{V}},
\end{align}
with $\omega_r$ the resonance frequency and $g_3$ the nonlinear coupling as in the main text.
We know from Eq.~\eqref{eq-app:S1_generator} that $\hat{S}^{(1)}$ will be given by
\begin{align}
    \hat{S}^{(1)} = \frac{g_3}{\omega_r} \left( \frac{1}{3} \hat{a}^{\dagger 3} + 3 \hat{a}^{\dagger} \hat{a} \hat{a}^{\dagger} - \mathrm{H.c.} \right),
\end{align}
which enables us to calculate $\hat{V}^{(2)}$ from Eq.~\eqref{eq:schrieffer_wolff_series_expansion}.
Furthermore, since the commutator $\comm*{\hat{S}^{(2)}}{\hat{H}_0}$ will only affect the off-diagonal terms of $\hat{V}^{(2)}$, one can obtain $\hat{H}^{(2)}$ from the diagonal part of $\hat{V}^{(2)}$.
For the example considered here, we find that $\hat{H}^{(2)}$ is given by
\begin{align}
    \hat{H}^{(2)} = \frac{g_3^2}{3 \omega_r} \comm{\hat{a}^{\dagger 3}}{\hat{a}^{3}} + 9 \frac{g_3^2}{\omega_r} \comm{\hat{a}^{\dagger} \hat{a} \hat{a}^{\dagger}}{\hat{a} \hat{a}^{\dagger} \hat{a}},
\end{align}
which is reminiscent of Eq.~\eqref{eq:H_eff_short}.

In the following we say that an operator $\hat{O}$ is normally ordered if all annihilation operators $\hat{a}$ stand to the right of the creation operators $\hat{a}^{\dagger}$.
The normal ordering should be distinguished from the double dot operation $:\hat{O}:$ which also moves all annihilation operators to right, but without taking commutation relations into account, i.e., operators are treated as numbers.
A general boson expression can be normal ordered by using Wick's theorem~\cite{wick_evaluation_1950, blasiak_combinatorics_2007} which relates the normal ordering to the double dot operation.
In particular, one obtains the normal ordered expression by applying the double dot operation to all possible expressions obtained by removing pairs of annihilation and creation operators for which $\hat{a}$ stands left of $\hat{a}^{\dagger}$ and summing over all contributions~\cite{blasiak_combinatorics_2007}.
Removing these pairs is known as performing contractions and is denoted with a square bracket over the operators which are being contracted, e.g.,
\begin{align}
    \hat{a}^{\dagger} \hat{a} \hat{a}^{\dagger} \hat{a} \hat{a}^{\dagger} \hat{a} =& :\hat{a}^{\dagger} \hat{a} \hat{a}^{\dagger} \hat{a} \hat{a}^{\dagger} \hat{a}: \nonumber \\
    &+ :\wick{\hat{a}^{\dagger} \c{\hat{a}} \c{\hat{a}}^{\dagger} \hat{a} \hat{a}^{\dagger} \hat{a}}  + \wick{\hat{a}^{\dagger} \c{\hat{a}} \hat{a}^{\dagger} \hat{a} \c{\hat{a}}^{\dagger} \hat{a}} + \wick{\hat{a}^{\dagger} \hat{a} \hat{a}^{\dagger} \c{\hat{a}} \c{\hat{a}}^{\dagger} \hat{a}}: \nonumber \\
    &+ :\hat{a}^{\dagger} \wick{\c{\hat{a}} \c{\hat{a}}^{\dagger}} \wick{\c{\hat{a}} \c{\hat{a}}^{\dagger} \hat{a}}: \nonumber \\
    &= \hat{a}^{\dagger 3} \hat{a}^{3} + 3 \hat{a}^{\dagger 2} \hat{a}^{2} + \hat{a}^{\dagger} \hat{a}, \label{eq-app:normal_order_example}
\end{align}
where in the first line no pairs are contracted, in the second line always one pair is contracted, and in the third line two pairs are contracted.
Notice also that the above example corresponds to the first terms of the commutator $\comm*{\hat{a}^{\dagger} \hat{a} \hat{a}^{\dagger}}{\hat{a} \hat{a}^{\dagger} \hat{a}}$.

It is an useful exercise to perform the normal ordering for the second term of the commutator $\hat{a} \hat{a}^{\dagger} \hat{a} \hat{a}^{\dagger} \hat{a} \hat{a}^{\dagger}$ as well and subtract it from the expression in Eq.~\eqref{eq-app:normal_order_example}.
One makes the following observations: (i) The term that does not contain any contraction will be canceled and
(ii) the only terms that contribute to the commutator are the ones where the contraction connects one of the first three operators with one of the last three ones.
In our qualitative diagrammatic picture, observation (i) leads to the ``rule'' that at order $m$ there are at least $m-1$ Wick contractions and observation (ii) exemplifies that only connected diagrams contribute.  
Furthermore, the scaling of the prefactor at $m$-th order, here $g_3^{m} / \omega_r^{m-1}$, follows from the iterative nature of Eq.~\eqref{eq:schrieffer_wolff_series_expansion} and the formal solution Eq.~\eqref{eq:SW_Sm_ansatz}.

\begin{figure}[!hb]
    \centering
    \includegraphics{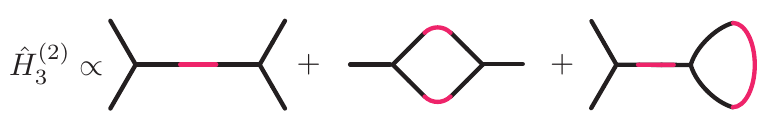}
    \caption{Diagrammatic representation of $\hat{H}_{3}^{(2)}$. The first diagram gives a correction to the Kerr term $(\hat{a}^{\dagger 2}\hat{a}^2)$ as it has four uncontracted legs. The last two diagrams have two uncontracted legs and yield corrections to the frequency. The placement and number of Wick contractions (red lines) agrees with the contractions that occur during normal ordering, cf. Eq.~\eqref{eq-app:normal_order_example}. We have omitted a diagram that is proportional to the identify , i.e., all legs are contracted, as it does not contribute to the dynamics. }
    \label{fig:app_diag_rep}
\end{figure}

We leave the task of formalizing this intuitive picture and deriving combinatorial rules for the numerical prefactors for future work.
By calculating the relevant commutators explicitly one finds that the correction for the Kerr term is given by $K = - 60 g_3^2 / \omega_r$ and the correction for detuning is given by $\delta = -60 g_3^2 / \omega_r$. 
All processes that contribute to these corrections can be visualized diagrammatically within our qualitative picture, see \figref{fig:app_diag_rep}.
The diagrams for higher-order perturbation theory are analogously constructed.

\section{Oscillation Average Procedure \label{app:oscillation_average}}

In the main text we have used an average over the fast oscillations when seeking the optimal value of $g_3$ to cancel self- or cross-Kerr effects.
This is done to obtain a quantity from the expectation value $\lvert \ev{\hat{a}} \rvert$ ($P(\hat{\rho}_a)$) that is independent of the fast oscillations induced by the off-resonant terms.
Although the averaging procedure can be performed in multiple ways, we found that a Savitzky-Golay filter~\cite{savitzky_smoothing_1964} implemented through {\small \texttt{scipy.signal.savgol\_filter}}~\cite{virtanen_scipy_2020} yielded the most stable results with respect to boundary effects.

\begin{figure}[H]
    \centering
    \includegraphics{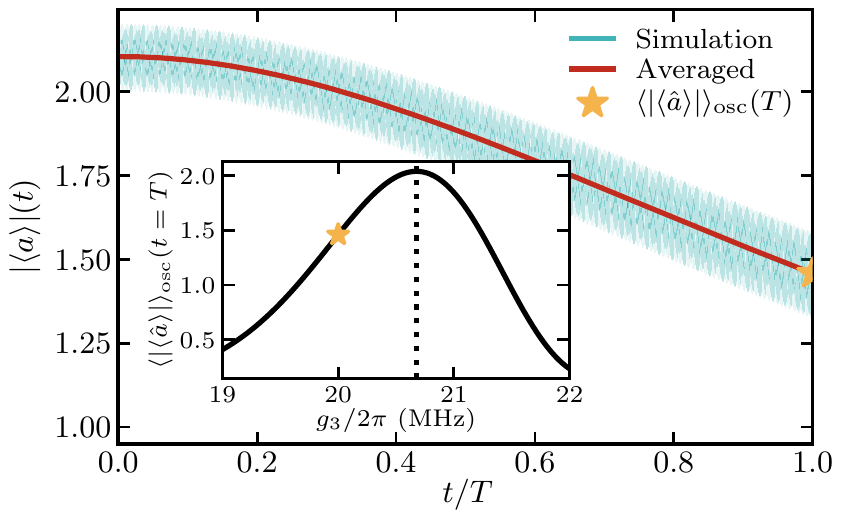}
    \caption{Visualization of the oscillation average procedure.
	Visualization of our methods described above for $(g_3 \dtwopi, g_4 \dtwopi) = (\simhz{20}, \simhz{0.5})$.
	Inset:~Value of $\langle \lvert \langle \hat{a} \rangle \rvert \rangle_{\mathrm{osc}}(t=T)$ for different values of $g_3$. The curve has a maximum for $\overline{g}_3 \dtwopi \approx \simhz{20.67}$ for which $\langle \lvert \langle \hat{a} \rangle \rvert \rangle_{\mathrm{osc}}(t=T) \approx 2$ indicates a large increase in the periodicity $T$ of Kerr oscillations.}
    \label{fig:osc_avg_proc}
\end{figure}

In \figref{fig:osc_avg_proc} the averaging procedure is exemplified for the simulations performed for \secref{ssec:num_self_kerr_renorm}.
All simulations are done using QuTiP~\cite{johansson_qutip_2013}.
Note that on this time-scale the fast oscillations are not resolved in the figure and appear (almost) as a thick blue line.
\\

%

%


\begin{thebibliography}{77}%
\makeatletter
\providecommand \@ifxundefined [1]{%
 \@ifx{#1\undefined}
}%
\providecommand \@ifnum [1]{%
 \ifnum #1\expandafter \@firstoftwo
 \else \expandafter \@secondoftwo
 \fi
}%
\providecommand \@ifx [1]{%
 \ifx #1\expandafter \@firstoftwo
 \else \expandafter \@secondoftwo
 \fi
}%
\providecommand \natexlab [1]{#1}%
\providecommand \enquote  [1]{``#1''}%
\providecommand \bibnamefont  [1]{#1}%
\providecommand \bibfnamefont [1]{#1}%
\providecommand \citenamefont [1]{#1}%
\providecommand \href@noop [0]{\@secondoftwo}%
\providecommand \href [0]{\begingroup \@sanitize@url \@href}%
\providecommand \@href[1]{\@@startlink{#1}\@@href}%
\providecommand \@@href[1]{\endgroup#1\@@endlink}%
\providecommand \@sanitize@url [0]{\catcode `\\12\catcode `\$12\catcode
  `\&12\catcode `\#12\catcode `\^12\catcode `\_12\catcode `\%12\relax}%
\providecommand \@@startlink[1]{}%
\providecommand \@@endlink[0]{}%
\providecommand \url  [0]{\begingroup\@sanitize@url \@url }%
\providecommand \@url [1]{\endgroup\@href {#1}{\urlprefix }}%
\providecommand \urlprefix  [0]{URL }%
\providecommand \Eprint [0]{\href }%
\providecommand \doibase [0]{https://doi.org/}%
\providecommand \selectlanguage [0]{\@gobble}%
\providecommand \bibinfo  [0]{\@secondoftwo}%
\providecommand \bibfield  [0]{\@secondoftwo}%
\providecommand \translation [1]{[#1]}%
\providecommand \BibitemOpen [0]{}%
\providecommand \bibitemStop [0]{}%
\providecommand \bibitemNoStop [0]{.\EOS\space}%
\providecommand \EOS [0]{\spacefactor3000\relax}%
\providecommand \BibitemShut  [1]{\csname bibitem#1\endcsname}%
\let\auto@bib@innerbib\@empty
\bibitem [{\citenamefont {Bruzewicz}\ \emph {et~al.}(2019)\citenamefont
  {Bruzewicz}, \citenamefont {Chiaverini}, \citenamefont {McConnell},\ and\
  \citenamefont {Sage}}]{bruzewicz_trapped-ion_2019}%
  \BibitemOpen
  \bibfield  {author} {\bibinfo {author} {\bibfnamefont {C.~D.}\ \bibnamefont
  {Bruzewicz}}, \bibinfo {author} {\bibfnamefont {J.}~\bibnamefont
  {Chiaverini}}, \bibinfo {author} {\bibfnamefont {R.}~\bibnamefont
  {McConnell}},\ and\ \bibinfo {author} {\bibfnamefont {J.~M.}\ \bibnamefont
  {Sage}},\ }\bibfield  {title} {\emph {\bibinfo {title} {Trapped-ion quantum
  computing: {Progress} and challenges}},\ }\href
  {https://doi.org/10.1063/1.5088164} {\bibfield  {journal} {\bibinfo
  {journal} {Appl. Phys. Rev}\ }\textbf {\bibinfo {volume} {6}},\ \bibinfo
  {pages} {021314} (\bibinfo {year} {2019})}\BibitemShut {NoStop}%
\bibitem [{\citenamefont {Wineland}(2013)}]{wineland_nobel_2013}%
  \BibitemOpen
  \bibfield  {author} {\bibinfo {author} {\bibfnamefont {D.~J.}\ \bibnamefont
  {Wineland}},\ }\bibfield  {title} {\emph {\bibinfo {title} {Nobel {Lecture}:
  {Superposition}, entanglement, and raising {Schrödinger}’s cat}},\ }\href
  {https://doi.org/10.1103/RevModPhys.85.1103} {\bibfield  {journal} {\bibinfo
  {journal} {Rev. Mod. Phys.}\ }\textbf {\bibinfo {volume} {85}},\ \bibinfo
  {pages} {1103} (\bibinfo {year} {2013})}\BibitemShut {NoStop}%
\bibitem [{\citenamefont {Devoret}\ and\ \citenamefont
  {Schoelkopf}(2013)}]{devoret_superconducting_2013}%
  \BibitemOpen
  \bibfield  {author} {\bibinfo {author} {\bibfnamefont {M.~H.}\ \bibnamefont
  {Devoret}}\ and\ \bibinfo {author} {\bibfnamefont {R.~J.}\ \bibnamefont
  {Schoelkopf}},\ }\bibfield  {title} {\emph {\bibinfo {title} {Superconducting
  {Circuits} for {Quantum} {Information}: {An} {Outlook}}},\ }\href
  {https://doi.org/10.1126/science.1231930} {\bibfield  {journal} {\bibinfo
  {journal} {Science}\ }\textbf {\bibinfo {volume} {339}},\ \bibinfo {pages}
  {1169} (\bibinfo {year} {2013})}\BibitemShut {NoStop}%
\bibitem [{\citenamefont {Blais}\ \emph {et~al.}(2021)\citenamefont {Blais},
  \citenamefont {Grimsmo}, \citenamefont {Girvin},\ and\ \citenamefont
  {Wallraff}}]{blais_circuit_2021}%
  \BibitemOpen
  \bibfield  {author} {\bibinfo {author} {\bibfnamefont {A.}~\bibnamefont
  {Blais}}, \bibinfo {author} {\bibfnamefont {A.~L.}\ \bibnamefont {Grimsmo}},
  \bibinfo {author} {\bibfnamefont {S.}~\bibnamefont {Girvin}},\ and\ \bibinfo
  {author} {\bibfnamefont {A.}~\bibnamefont {Wallraff}},\ }\bibfield  {title}
  {\emph {\bibinfo {title} {Circuit quantum electrodynamics}},\ }\href
  {https://doi.org/10.1103/RevModPhys.93.025005} {\bibfield  {journal}
  {\bibinfo  {journal} {Rev. Mod. Phys.}\ }\textbf {\bibinfo {volume} {93}},\
  \bibinfo {pages} {025005} (\bibinfo {year} {2021})}\BibitemShut {NoStop}%
\bibitem [{\citenamefont {Kloeffel}\ and\ \citenamefont
  {Loss}(2013)}]{kloeffel_prospects_2013}%
  \BibitemOpen
  \bibfield  {author} {\bibinfo {author} {\bibfnamefont {C.}~\bibnamefont
  {Kloeffel}}\ and\ \bibinfo {author} {\bibfnamefont {D.}~\bibnamefont
  {Loss}},\ }\bibfield  {title} {\emph {\bibinfo {title} {Prospects for
  {Spin}-{Based} {Quantum} {Computing} in {Quantum} {Dots}}},\ }\href
  {https://doi.org/10.1146/annurev-conmatphys-030212-184248} {\bibfield
  {journal} {\bibinfo  {journal} {Annu. Rev. Condens. Matter Phys.}\ }\textbf
  {\bibinfo {volume} {4}},\ \bibinfo {pages} {51} (\bibinfo {year}
  {2013})}\BibitemShut {NoStop}%
\bibitem [{\citenamefont {Ofek}\ \emph {et~al.}(2016)\citenamefont {Ofek},
  \citenamefont {Petrenko}, \citenamefont {Heeres}, \citenamefont {Reinhold},
  \citenamefont {Leghtas}, \citenamefont {Vlastakis}, \citenamefont {Liu},
  \citenamefont {Frunzio}, \citenamefont {Girvin}, \citenamefont {Jiang},
  \citenamefont {Mirrahimi}, \citenamefont {Devoret},\ and\ \citenamefont
  {Schoelkopf}}]{ofek_extending_2016}%
  \BibitemOpen
  \bibfield  {author} {\bibinfo {author} {\bibfnamefont {N.}~\bibnamefont
  {Ofek}}, \bibinfo {author} {\bibfnamefont {A.}~\bibnamefont {Petrenko}},
  \bibinfo {author} {\bibfnamefont {R.}~\bibnamefont {Heeres}}, \bibinfo
  {author} {\bibfnamefont {P.}~\bibnamefont {Reinhold}}, \bibinfo {author}
  {\bibfnamefont {Z.}~\bibnamefont {Leghtas}}, \bibinfo {author} {\bibfnamefont
  {B.}~\bibnamefont {Vlastakis}}, \bibinfo {author} {\bibfnamefont
  {Y.}~\bibnamefont {Liu}}, \bibinfo {author} {\bibfnamefont {L.}~\bibnamefont
  {Frunzio}}, \bibinfo {author} {\bibfnamefont {S.~M.}\ \bibnamefont {Girvin}},
  \bibinfo {author} {\bibfnamefont {L.}~\bibnamefont {Jiang}}, \bibinfo
  {author} {\bibfnamefont {M.}~\bibnamefont {Mirrahimi}}, \bibinfo {author}
  {\bibfnamefont {M.~H.}\ \bibnamefont {Devoret}},\ and\ \bibinfo {author}
  {\bibfnamefont {R.~J.}\ \bibnamefont {Schoelkopf}},\ }\bibfield  {title}
  {\emph {\bibinfo {title} {Extending the lifetime of a quantum bit with error
  correction in superconducting circuits}},\ }\href
  {https://doi.org/10.1038/nature18949} {\bibfield  {journal} {\bibinfo
  {journal} {Nature}\ }\textbf {\bibinfo {volume} {536}},\ \bibinfo {pages}
  {441} (\bibinfo {year} {2016})}\BibitemShut {NoStop}%
\bibitem [{\citenamefont {Hu}\ \emph {et~al.}(2019)\citenamefont {Hu},
  \citenamefont {Ma}, \citenamefont {Cai}, \citenamefont {Mu}, \citenamefont
  {Xu}, \citenamefont {Wang}, \citenamefont {Wu}, \citenamefont {Wang},
  \citenamefont {Song}, \citenamefont {Zou}, \citenamefont {Girvin},
  \citenamefont {Duan},\ and\ \citenamefont {Sun}}]{hu_quantum_2019}%
  \BibitemOpen
  \bibfield  {author} {\bibinfo {author} {\bibfnamefont {L.}~\bibnamefont
  {Hu}}, \bibinfo {author} {\bibfnamefont {Y.}~\bibnamefont {Ma}}, \bibinfo
  {author} {\bibfnamefont {W.}~\bibnamefont {Cai}}, \bibinfo {author}
  {\bibfnamefont {X.}~\bibnamefont {Mu}}, \bibinfo {author} {\bibfnamefont
  {Y.}~\bibnamefont {Xu}}, \bibinfo {author} {\bibfnamefont {W.}~\bibnamefont
  {Wang}}, \bibinfo {author} {\bibfnamefont {Y.}~\bibnamefont {Wu}}, \bibinfo
  {author} {\bibfnamefont {H.}~\bibnamefont {Wang}}, \bibinfo {author}
  {\bibfnamefont {Y.~P.}\ \bibnamefont {Song}}, \bibinfo {author}
  {\bibfnamefont {C.-L.}\ \bibnamefont {Zou}}, \bibinfo {author} {\bibfnamefont
  {S.~M.}\ \bibnamefont {Girvin}}, \bibinfo {author} {\bibfnamefont {L.-M.}\
  \bibnamefont {Duan}},\ and\ \bibinfo {author} {\bibfnamefont
  {L.}~\bibnamefont {Sun}},\ }\bibfield  {title} {\emph {\bibinfo {title}
  {Quantum error correction and universal gate set operation on a binomial
  bosonic logical qubit}},\ }\href {https://doi.org/10.1038/s41567-018-0414-3}
  {\bibfield  {journal} {\bibinfo  {journal} {Nat. Phys.}\ }\textbf {\bibinfo
  {volume} {15}},\ \bibinfo {pages} {503} (\bibinfo {year} {2019})}\BibitemShut
  {NoStop}%
\bibitem [{\citenamefont {Gertler}\ \emph {et~al.}(2021)\citenamefont
  {Gertler}, \citenamefont {Baker}, \citenamefont {Li}, \citenamefont {Shirol},
  \citenamefont {Koch},\ and\ \citenamefont {Wang}}]{gertler_protecting_2021}%
  \BibitemOpen
  \bibfield  {author} {\bibinfo {author} {\bibfnamefont {J.~M.}\ \bibnamefont
  {Gertler}}, \bibinfo {author} {\bibfnamefont {B.}~\bibnamefont {Baker}},
  \bibinfo {author} {\bibfnamefont {J.}~\bibnamefont {Li}}, \bibinfo {author}
  {\bibfnamefont {S.}~\bibnamefont {Shirol}}, \bibinfo {author} {\bibfnamefont
  {J.}~\bibnamefont {Koch}},\ and\ \bibinfo {author} {\bibfnamefont
  {C.}~\bibnamefont {Wang}},\ }\bibfield  {title} {\emph {\bibinfo {title}
  {Protecting a bosonic qubit with autonomous quantum error correction}},\
  }\href {https://doi.org/10.1038/s41586-021-03257-0} {\bibfield  {journal}
  {\bibinfo  {journal} {Nature}\ }\textbf {\bibinfo {volume} {590}},\ \bibinfo
  {pages} {243} (\bibinfo {year} {2021})}\BibitemShut {NoStop}%
\bibitem [{\citenamefont {Haroche}(2013)}]{haroche_nobel_2013}%
  \BibitemOpen
  \bibfield  {author} {\bibinfo {author} {\bibfnamefont {S.}~\bibnamefont
  {Haroche}},\ }\bibfield  {title} {\emph {\bibinfo {title} {Nobel {Lecture}:
  {Controlling} photons in a box and exploring the quantum to classical
  boundary}},\ }\href {https://doi.org/10.1103/RevModPhys.85.1083} {\bibfield
  {journal} {\bibinfo  {journal} {Rev. Mod. Phys.}\ }\textbf {\bibinfo {volume}
  {85}},\ \bibinfo {pages} {1083} (\bibinfo {year} {2013})}\BibitemShut
  {NoStop}%
\bibitem [{\citenamefont {Aspelmeyer}\ \emph {et~al.}(2014)\citenamefont
  {Aspelmeyer}, \citenamefont {Kippenberg},\ and\ \citenamefont
  {Marquardt}}]{aspelmeyer_cavity_2014}%
  \BibitemOpen
  \bibfield  {author} {\bibinfo {author} {\bibfnamefont {M.}~\bibnamefont
  {Aspelmeyer}}, \bibinfo {author} {\bibfnamefont {T.~J.}\ \bibnamefont
  {Kippenberg}},\ and\ \bibinfo {author} {\bibfnamefont {F.}~\bibnamefont
  {Marquardt}},\ }\bibfield  {title} {\emph {\bibinfo {title} {Cavity
  optomechanics}},\ }\href {https://doi.org/10.1103/RevModPhys.86.1391}
  {\bibfield  {journal} {\bibinfo  {journal} {Rev. Mod. Phys.}\ }\textbf
  {\bibinfo {volume} {86}},\ \bibinfo {pages} {1391} (\bibinfo {year}
  {2014})}\BibitemShut {NoStop}%
\bibitem [{\citenamefont {Joshi}\ \emph {et~al.}(2020)\citenamefont {Joshi},
  \citenamefont {Noh},\ and\ \citenamefont {Gao}}]{joshi_quantum_2020}%
  \BibitemOpen
  \bibfield  {author} {\bibinfo {author} {\bibfnamefont {A.}~\bibnamefont
  {Joshi}}, \bibinfo {author} {\bibfnamefont {K.}~\bibnamefont {Noh}},\ and\
  \bibinfo {author} {\bibfnamefont {Y.~Y.}\ \bibnamefont {Gao}},\ }\bibfield
  {title} {\emph {\bibinfo {title} {Quantum information processing with bosonic
  qubits in circuit {QED}}},\ }\href {http://arxiv.org/abs/2008.13471}
  {\bibfield  {journal} {\bibinfo  {journal} {arXiv:2008.13471 [quant-ph]}\ }
  (\bibinfo {year} {2020})}\BibitemShut {NoStop}%
\bibitem [{\citenamefont {Romanenko}\ \emph {et~al.}(2020)\citenamefont
  {Romanenko}, \citenamefont {Pilipenko}, \citenamefont {Zorzetti},
  \citenamefont {Frolov}, \citenamefont {Awida}, \citenamefont {Belomestnykh},
  \citenamefont {Posen},\ and\ \citenamefont
  {Grassellino}}]{romanenko_three-dimensional_2020}%
  \BibitemOpen
  \bibfield  {author} {\bibinfo {author} {\bibfnamefont {A.}~\bibnamefont
  {Romanenko}}, \bibinfo {author} {\bibfnamefont {R.}~\bibnamefont
  {Pilipenko}}, \bibinfo {author} {\bibfnamefont {S.}~\bibnamefont {Zorzetti}},
  \bibinfo {author} {\bibfnamefont {D.}~\bibnamefont {Frolov}}, \bibinfo
  {author} {\bibfnamefont {M.}~\bibnamefont {Awida}}, \bibinfo {author}
  {\bibfnamefont {S.}~\bibnamefont {Belomestnykh}}, \bibinfo {author}
  {\bibfnamefont {S.}~\bibnamefont {Posen}},\ and\ \bibinfo {author}
  {\bibfnamefont {A.}~\bibnamefont {Grassellino}},\ }\bibfield  {title} {\emph
  {\bibinfo {title} {Three-{Dimensional} {Superconducting} {Resonators} at {$T
  < 20$} {mK} with {Photon} {Lifetimes} up to {$\tau=2$} s}},\ }\href
  {https://doi.org/10.1103/PhysRevApplied.13.034032} {\bibfield  {journal}
  {\bibinfo  {journal} {Phys. Rev. Applied}\ }\textbf {\bibinfo {volume}
  {13}},\ \bibinfo {pages} {034032} (\bibinfo {year} {2020})}\BibitemShut
  {NoStop}%
\bibitem [{\citenamefont {Kudra}\ \emph {et~al.}(2020)\citenamefont {Kudra},
  \citenamefont {Biznárová}, \citenamefont {Fadavi~Roudsari}, \citenamefont
  {Burnett}, \citenamefont {Niepce}, \citenamefont {Gasparinetti},
  \citenamefont {Wickman},\ and\ \citenamefont {Delsing}}]{kudra_high_2020}%
  \BibitemOpen
  \bibfield  {author} {\bibinfo {author} {\bibfnamefont {M.}~\bibnamefont
  {Kudra}}, \bibinfo {author} {\bibfnamefont {J.}~\bibnamefont {Biznárová}},
  \bibinfo {author} {\bibfnamefont {A.}~\bibnamefont {Fadavi~Roudsari}},
  \bibinfo {author} {\bibfnamefont {J.~J.}\ \bibnamefont {Burnett}}, \bibinfo
  {author} {\bibfnamefont {D.}~\bibnamefont {Niepce}}, \bibinfo {author}
  {\bibfnamefont {S.}~\bibnamefont {Gasparinetti}}, \bibinfo {author}
  {\bibfnamefont {B.}~\bibnamefont {Wickman}},\ and\ \bibinfo {author}
  {\bibfnamefont {P.}~\bibnamefont {Delsing}},\ }\bibfield  {title} {\emph
  {\bibinfo {title} {High quality three-dimensional aluminum microwave
  cavities}},\ }\href {https://doi.org/10.1063/5.0016463} {\bibfield  {journal}
  {\bibinfo  {journal} {Appl. Phys. Lett.}\ }\textbf {\bibinfo {volume}
  {117}},\ \bibinfo {pages} {070601} (\bibinfo {year} {2020})}\BibitemShut
  {NoStop}%
\bibitem [{\citenamefont {Heidler}\ \emph {et~al.}(2021)\citenamefont
  {Heidler}, \citenamefont {Schneider}, \citenamefont {Kustura}, \citenamefont
  {Gonzalez-Ballestero}, \citenamefont {Romero-Isart},\ and\ \citenamefont
  {Kirchmair}}]{heidler_observing_2021}%
  \BibitemOpen
  \bibfield  {author} {\bibinfo {author} {\bibfnamefont {P.}~\bibnamefont
  {Heidler}}, \bibinfo {author} {\bibfnamefont {C.~M.~F.}\ \bibnamefont
  {Schneider}}, \bibinfo {author} {\bibfnamefont {K.}~\bibnamefont {Kustura}},
  \bibinfo {author} {\bibfnamefont {C.}~\bibnamefont {Gonzalez-Ballestero}},
  \bibinfo {author} {\bibfnamefont {O.}~\bibnamefont {Romero-Isart}},\ and\
  \bibinfo {author} {\bibfnamefont {G.}~\bibnamefont {Kirchmair}},\ }\bibfield
  {title} {\emph {\bibinfo {title} {Observing {Non}-{Markovian} {Effects} of
  {Two}-{Level} {Systems} in a {Niobium} {Coaxial} {Resonator} with a
  {Single}-{Photon} {Lifetime} of 10 ms}},\ }\href
  {http://arxiv.org/abs/2102.10016} {\bibfield  {journal} {\bibinfo  {journal}
  {arXiv:2102.10016 [cond-mat, physics:quant-ph]}\ } (\bibinfo {year}
  {2021})}\BibitemShut {NoStop}%
\bibitem [{\citenamefont {Axline}\ \emph {et~al.}(2016)\citenamefont {Axline},
  \citenamefont {Reagor}, \citenamefont {Heeres}, \citenamefont {Reinhold},
  \citenamefont {Wang}, \citenamefont {Shain}, \citenamefont {Pfaff},
  \citenamefont {Chu}, \citenamefont {Frunzio},\ and\ \citenamefont
  {Schoelkopf}}]{axline_architecture_2016}%
  \BibitemOpen
  \bibfield  {author} {\bibinfo {author} {\bibfnamefont {C.}~\bibnamefont
  {Axline}}, \bibinfo {author} {\bibfnamefont {M.}~\bibnamefont {Reagor}},
  \bibinfo {author} {\bibfnamefont {R.}~\bibnamefont {Heeres}}, \bibinfo
  {author} {\bibfnamefont {P.}~\bibnamefont {Reinhold}}, \bibinfo {author}
  {\bibfnamefont {C.}~\bibnamefont {Wang}}, \bibinfo {author} {\bibfnamefont
  {K.}~\bibnamefont {Shain}}, \bibinfo {author} {\bibfnamefont
  {W.}~\bibnamefont {Pfaff}}, \bibinfo {author} {\bibfnamefont
  {Y.}~\bibnamefont {Chu}}, \bibinfo {author} {\bibfnamefont {L.}~\bibnamefont
  {Frunzio}},\ and\ \bibinfo {author} {\bibfnamefont {R.~J.}\ \bibnamefont
  {Schoelkopf}},\ }\bibfield  {title} {\emph {\bibinfo {title} {An architecture
  for integrating planar and {3D} {cQED} devices}},\ }\href
  {https://doi.org/10.1063/1.4959241} {\bibfield  {journal} {\bibinfo
  {journal} {Appl. Phys. Lett.}\ }\textbf {\bibinfo {volume} {109}},\ \bibinfo
  {pages} {042601} (\bibinfo {year} {2016})}\BibitemShut {NoStop}%
\bibitem [{\citenamefont {Lloyd}\ and\ \citenamefont
  {Braunstein}(1999)}]{lloyd_quantum_1999}%
  \BibitemOpen
  \bibfield  {author} {\bibinfo {author} {\bibfnamefont {S.}~\bibnamefont
  {Lloyd}}\ and\ \bibinfo {author} {\bibfnamefont {S.~L.}\ \bibnamefont
  {Braunstein}},\ }\bibfield  {title} {\emph {\bibinfo {title} {Quantum
  {Computation} over {Continuous} {Variables}}},\ }\href
  {https://doi.org/10.1103/PhysRevLett.82.1784} {\bibfield  {journal} {\bibinfo
   {journal} {Phys. Rev. Lett.}\ }\textbf {\bibinfo {volume} {82}},\ \bibinfo
  {pages} {1784} (\bibinfo {year} {1999})}\BibitemShut {NoStop}%
\bibitem [{\citenamefont {Campagne-Ibarcq}\ \emph {et~al.}(2020)\citenamefont
  {Campagne-Ibarcq}, \citenamefont {Eickbusch}, \citenamefont {Touzard},
  \citenamefont {Zalys-Geller}, \citenamefont {Frattini}, \citenamefont
  {Sivak}, \citenamefont {Reinhold}, \citenamefont {Puri}, \citenamefont
  {Shankar}, \citenamefont {Schoelkopf}, \citenamefont {Frunzio}, \citenamefont
  {Mirrahimi},\ and\ \citenamefont {Devoret}}]{campagne-ibarcq_quantum_2020}%
  \BibitemOpen
  \bibfield  {author} {\bibinfo {author} {\bibfnamefont {P.}~\bibnamefont
  {Campagne-Ibarcq}}, \bibinfo {author} {\bibfnamefont {A.}~\bibnamefont
  {Eickbusch}}, \bibinfo {author} {\bibfnamefont {S.}~\bibnamefont {Touzard}},
  \bibinfo {author} {\bibfnamefont {E.}~\bibnamefont {Zalys-Geller}}, \bibinfo
  {author} {\bibfnamefont {N.~E.}\ \bibnamefont {Frattini}}, \bibinfo {author}
  {\bibfnamefont {V.~V.}\ \bibnamefont {Sivak}}, \bibinfo {author}
  {\bibfnamefont {P.}~\bibnamefont {Reinhold}}, \bibinfo {author}
  {\bibfnamefont {S.}~\bibnamefont {Puri}}, \bibinfo {author} {\bibfnamefont
  {S.}~\bibnamefont {Shankar}}, \bibinfo {author} {\bibfnamefont {R.~J.}\
  \bibnamefont {Schoelkopf}}, \bibinfo {author} {\bibfnamefont
  {L.}~\bibnamefont {Frunzio}}, \bibinfo {author} {\bibfnamefont
  {M.}~\bibnamefont {Mirrahimi}},\ and\ \bibinfo {author} {\bibfnamefont
  {M.~H.}\ \bibnamefont {Devoret}},\ }\bibfield  {title} {\emph {\bibinfo
  {title} {Quantum error correction of a qubit encoded in grid states of an
  oscillator}},\ }\href {https://doi.org/10.1038/s41586-020-2603-3} {\bibfield
  {journal} {\bibinfo  {journal} {Nature}\ }\textbf {\bibinfo {volume} {584}},\
  \bibinfo {pages} {368} (\bibinfo {year} {2020})}\BibitemShut {NoStop}%
\bibitem [{\citenamefont {Ma}\ \emph {et~al.}(2020)\citenamefont {Ma},
  \citenamefont {Xu}, \citenamefont {Mu}, \citenamefont {Cai}, \citenamefont
  {Hu}, \citenamefont {Wang}, \citenamefont {Pan}, \citenamefont {Wang},
  \citenamefont {Song}, \citenamefont {Zou},\ and\ \citenamefont
  {Sun}}]{ma_error-transparent_2020}%
  \BibitemOpen
  \bibfield  {author} {\bibinfo {author} {\bibfnamefont {Y.}~\bibnamefont
  {Ma}}, \bibinfo {author} {\bibfnamefont {Y.}~\bibnamefont {Xu}}, \bibinfo
  {author} {\bibfnamefont {X.}~\bibnamefont {Mu}}, \bibinfo {author}
  {\bibfnamefont {W.}~\bibnamefont {Cai}}, \bibinfo {author} {\bibfnamefont
  {L.}~\bibnamefont {Hu}}, \bibinfo {author} {\bibfnamefont {W.}~\bibnamefont
  {Wang}}, \bibinfo {author} {\bibfnamefont {X.}~\bibnamefont {Pan}}, \bibinfo
  {author} {\bibfnamefont {H.}~\bibnamefont {Wang}}, \bibinfo {author}
  {\bibfnamefont {Y.~P.}\ \bibnamefont {Song}}, \bibinfo {author}
  {\bibfnamefont {C.-L.}\ \bibnamefont {Zou}},\ and\ \bibinfo {author}
  {\bibfnamefont {L.}~\bibnamefont {Sun}},\ }\bibfield  {title} {\emph
  {\bibinfo {title} {Error-transparent operations on a logical qubit protected
  by quantum error correction}},\ }\href
  {https://doi.org/10.1038/s41567-020-0893-x} {\bibfield  {journal} {\bibinfo
  {journal} {Nat. Phys.}\ }\textbf {\bibinfo {volume} {16}},\ \bibinfo {pages}
  {827} (\bibinfo {year} {2020})}\BibitemShut {NoStop}%
\bibitem [{\citenamefont {Pfaff}\ \emph {et~al.}(2017)\citenamefont {Pfaff},
  \citenamefont {Axline}, \citenamefont {Burkhart}, \citenamefont {Vool},
  \citenamefont {Reinhold}, \citenamefont {Frunzio}, \citenamefont {Jiang},
  \citenamefont {Devoret},\ and\ \citenamefont
  {Schoelkopf}}]{pfaff_controlled_2017}%
  \BibitemOpen
  \bibfield  {author} {\bibinfo {author} {\bibfnamefont {W.}~\bibnamefont
  {Pfaff}}, \bibinfo {author} {\bibfnamefont {C.~J.}\ \bibnamefont {Axline}},
  \bibinfo {author} {\bibfnamefont {L.~D.}\ \bibnamefont {Burkhart}}, \bibinfo
  {author} {\bibfnamefont {U.}~\bibnamefont {Vool}}, \bibinfo {author}
  {\bibfnamefont {P.}~\bibnamefont {Reinhold}}, \bibinfo {author}
  {\bibfnamefont {L.}~\bibnamefont {Frunzio}}, \bibinfo {author} {\bibfnamefont
  {L.}~\bibnamefont {Jiang}}, \bibinfo {author} {\bibfnamefont {M.~H.}\
  \bibnamefont {Devoret}},\ and\ \bibinfo {author} {\bibfnamefont {R.~J.}\
  \bibnamefont {Schoelkopf}},\ }\bibfield  {title} {\emph {\bibinfo {title}
  {Controlled release of multiphoton quantum states from a microwave cavity
  memory}},\ }\href {https://doi.org/10.1038/nphys4143} {\bibfield  {journal}
  {\bibinfo  {journal} {Nat. Phys.}\ }\textbf {\bibinfo {volume} {13}},\
  \bibinfo {pages} {882} (\bibinfo {year} {2017})}\BibitemShut {NoStop}%
\bibitem [{\citenamefont {Lescanne}\ \emph
  {et~al.}(2020{\natexlab{a}})\citenamefont {Lescanne}, \citenamefont
  {Deléglise}, \citenamefont {Albertinale}, \citenamefont {Réglade},
  \citenamefont {Capelle}, \citenamefont {Ivanov}, \citenamefont {Jacqmin},
  \citenamefont {Leghtas},\ and\ \citenamefont
  {Flurin}}]{lescanne_irreversible_2020}%
  \BibitemOpen
  \bibfield  {author} {\bibinfo {author} {\bibfnamefont {R.}~\bibnamefont
  {Lescanne}}, \bibinfo {author} {\bibfnamefont {S.}~\bibnamefont
  {Deléglise}}, \bibinfo {author} {\bibfnamefont {E.}~\bibnamefont
  {Albertinale}}, \bibinfo {author} {\bibfnamefont {U.}~\bibnamefont
  {Réglade}}, \bibinfo {author} {\bibfnamefont {T.}~\bibnamefont {Capelle}},
  \bibinfo {author} {\bibfnamefont {E.}~\bibnamefont {Ivanov}}, \bibinfo
  {author} {\bibfnamefont {T.}~\bibnamefont {Jacqmin}}, \bibinfo {author}
  {\bibfnamefont {Z.}~\bibnamefont {Leghtas}},\ and\ \bibinfo {author}
  {\bibfnamefont {E.}~\bibnamefont {Flurin}},\ }\bibfield  {title} {\emph
  {\bibinfo {title} {Irreversible {Qubit}-{Photon} {Coupling} for the
  {Detection} of {Itinerant} {Microwave} {Photons}}},\ }\href
  {https://doi.org/10.1103/PhysRevX.10.021038} {\bibfield  {journal} {\bibinfo
  {journal} {Phys. Rev. X}\ }\textbf {\bibinfo {volume} {10}},\ \bibinfo
  {pages} {021038} (\bibinfo {year} {2020}{\natexlab{a}})}\BibitemShut
  {NoStop}%
\bibitem [{\citenamefont {Wang}\ \emph {et~al.}(2021)\citenamefont {Wang},
  \citenamefont {Noh}, \citenamefont {Lebreuilly}, \citenamefont {Girvin},\
  and\ \citenamefont {Jiang}}]{wang_photon-number-dependent_2021}%
  \BibitemOpen
  \bibfield  {author} {\bibinfo {author} {\bibfnamefont {C.-H.}\ \bibnamefont
  {Wang}}, \bibinfo {author} {\bibfnamefont {K.}~\bibnamefont {Noh}}, \bibinfo
  {author} {\bibfnamefont {J.}~\bibnamefont {Lebreuilly}}, \bibinfo {author}
  {\bibfnamefont {S.}~\bibnamefont {Girvin}},\ and\ \bibinfo {author}
  {\bibfnamefont {L.}~\bibnamefont {Jiang}},\ }\bibfield  {title} {\emph
  {\bibinfo {title} {Photon-{Number}-{Dependent} {Hamiltonian} {Engineering}
  for {Cavities}}},\ }\href {https://doi.org/10.1103/PhysRevApplied.15.044026}
  {\bibfield  {journal} {\bibinfo  {journal} {Phys. Rev. Applied}\ }\textbf
  {\bibinfo {volume} {15}},\ \bibinfo {pages} {044026} (\bibinfo {year}
  {2021})}\BibitemShut {NoStop}%
\bibitem [{\citenamefont {Zhang}\ \emph {et~al.}(2021)\citenamefont {Zhang},
  \citenamefont {Curtis}, \citenamefont {Wang}, \citenamefont {Schoelkopf},\
  and\ \citenamefont {Girvin}}]{zhang_drive-induced_2021}%
  \BibitemOpen
  \bibfield  {author} {\bibinfo {author} {\bibfnamefont {Y.}~\bibnamefont
  {Zhang}}, \bibinfo {author} {\bibfnamefont {J.~C.}\ \bibnamefont {Curtis}},
  \bibinfo {author} {\bibfnamefont {C.~S.}\ \bibnamefont {Wang}}, \bibinfo
  {author} {\bibfnamefont {R.~J.}\ \bibnamefont {Schoelkopf}},\ and\ \bibinfo
  {author} {\bibfnamefont {S.~M.}\ \bibnamefont {Girvin}},\ }\bibfield  {title}
  {\emph {\bibinfo {title} {Drive-induced nonlinearities of cavity modes
  coupled to a transmon ancilla}},\ }\href {http://arxiv.org/abs/2106.09112}
  {\bibfield  {journal} {\bibinfo  {journal} {arXiv:2106.09112 [cond-mat,
  physics:quant-ph]}\ } (\bibinfo {year} {2021})}\BibitemShut {NoStop}%
\bibitem [{\citenamefont {Petrescu}\ \emph
  {et~al.}(2021{\natexlab{a}})\citenamefont {Petrescu}, \citenamefont
  {Calonnec}, \citenamefont {Leroux}, \citenamefont {Di~Paolo}, \citenamefont
  {Mundada}, \citenamefont {Sussman}, \citenamefont {Vrajitoarea},
  \citenamefont {Houck},\ and\ \citenamefont
  {Blais}}]{petrescu_accurate_2021-1}%
  \BibitemOpen
  \bibfield  {author} {\bibinfo {author} {\bibfnamefont {A.}~\bibnamefont
  {Petrescu}}, \bibinfo {author} {\bibfnamefont {C.~L.}\ \bibnamefont
  {Calonnec}}, \bibinfo {author} {\bibfnamefont {C.}~\bibnamefont {Leroux}},
  \bibinfo {author} {\bibfnamefont {A.}~\bibnamefont {Di~Paolo}}, \bibinfo
  {author} {\bibfnamefont {P.}~\bibnamefont {Mundada}}, \bibinfo {author}
  {\bibfnamefont {S.}~\bibnamefont {Sussman}}, \bibinfo {author} {\bibfnamefont
  {A.}~\bibnamefont {Vrajitoarea}}, \bibinfo {author} {\bibfnamefont {A.~A.}\
  \bibnamefont {Houck}},\ and\ \bibinfo {author} {\bibfnamefont
  {A.}~\bibnamefont {Blais}},\ }\bibfield  {title} {\emph {\bibinfo {title}
  {Accurate methods for the analysis of strong-drive effects in parametric
  gates}},\ }\href {http://arxiv.org/abs/2107.02343} {\bibfield  {journal}
  {\bibinfo  {journal} {arXiv:2107.02343 [cond-mat, physics:quant-ph]}\ }
  (\bibinfo {year} {2021}{\natexlab{a}})}\BibitemShut {NoStop}%
\bibitem [{\citenamefont {Venkatraman}\ \emph {et~al.}(2021)\citenamefont
  {Venkatraman}, \citenamefont {Xiao}, \citenamefont {Corti{\~n}as},\ and\
  \citenamefont {Devoret}}]{venkatraman_static_2021}%
  \BibitemOpen
  \bibfield  {author} {\bibinfo {author} {\bibfnamefont {J.}~\bibnamefont
  {Venkatraman}}, \bibinfo {author} {\bibfnamefont {X.}~\bibnamefont {Xiao}},
  \bibinfo {author} {\bibfnamefont {R.~G.}\ \bibnamefont {Corti{\~n}as}},\ and\
  \bibinfo {author} {\bibfnamefont {M.~H.}\ \bibnamefont {Devoret}},\
  }\bibfield  {title} {\emph {\bibinfo {title} {On the static effective
  hamiltonian of a rapidly driven nonlinear system}},\ }\href
  {https://arxiv.org/abs/2108.02861} {\bibfield  {journal} {\bibinfo  {journal}
  {arXiv:2108.02861}\ } (\bibinfo {year} {2021})}\BibitemShut {NoStop}%
\bibitem [{\citenamefont {Frattini}\ \emph {et~al.}(2017)\citenamefont
  {Frattini}, \citenamefont {Vool}, \citenamefont {Shankar}, \citenamefont
  {Narla}, \citenamefont {Sliwa},\ and\ \citenamefont
  {Devoret}}]{frattini_3-wave_2017}%
  \BibitemOpen
  \bibfield  {author} {\bibinfo {author} {\bibfnamefont {N.~E.}\ \bibnamefont
  {Frattini}}, \bibinfo {author} {\bibfnamefont {U.}~\bibnamefont {Vool}},
  \bibinfo {author} {\bibfnamefont {S.}~\bibnamefont {Shankar}}, \bibinfo
  {author} {\bibfnamefont {A.}~\bibnamefont {Narla}}, \bibinfo {author}
  {\bibfnamefont {K.~M.}\ \bibnamefont {Sliwa}},\ and\ \bibinfo {author}
  {\bibfnamefont {M.~H.}\ \bibnamefont {Devoret}},\ }\bibfield  {title} {\emph
  {\bibinfo {title} {3-wave mixing {Josephson} dipole element}},\ }\href
  {https://doi.org/10.1063/1.4984142} {\bibfield  {journal} {\bibinfo
  {journal} {Appl. Phys. Lett.}\ }\textbf {\bibinfo {volume} {110}},\ \bibinfo
  {pages} {222603} (\bibinfo {year} {2017})}\BibitemShut {NoStop}%
\bibitem [{\citenamefont {Lescanne}\ \emph
  {et~al.}(2020{\natexlab{b}})\citenamefont {Lescanne}, \citenamefont
  {Villiers}, \citenamefont {Peronnin}, \citenamefont {Sarlette}, \citenamefont
  {Delbecq}, \citenamefont {Huard}, \citenamefont {Kontos}, \citenamefont
  {Mirrahimi},\ and\ \citenamefont {Leghtas}}]{lescanne_exponential_2020}%
  \BibitemOpen
  \bibfield  {author} {\bibinfo {author} {\bibfnamefont {R.}~\bibnamefont
  {Lescanne}}, \bibinfo {author} {\bibfnamefont {M.}~\bibnamefont {Villiers}},
  \bibinfo {author} {\bibfnamefont {T.}~\bibnamefont {Peronnin}}, \bibinfo
  {author} {\bibfnamefont {A.}~\bibnamefont {Sarlette}}, \bibinfo {author}
  {\bibfnamefont {M.}~\bibnamefont {Delbecq}}, \bibinfo {author} {\bibfnamefont
  {B.}~\bibnamefont {Huard}}, \bibinfo {author} {\bibfnamefont
  {T.}~\bibnamefont {Kontos}}, \bibinfo {author} {\bibfnamefont
  {M.}~\bibnamefont {Mirrahimi}},\ and\ \bibinfo {author} {\bibfnamefont
  {Z.}~\bibnamefont {Leghtas}},\ }\bibfield  {title} {\emph {\bibinfo {title}
  {Exponential suppression of bit-flips in a qubit encoded in an oscillator}},\
  }\href {https://doi.org/10.1038/s41567-020-0824-x} {\bibfield  {journal}
  {\bibinfo  {journal} {Nat. Phys.}\ }\textbf {\bibinfo {volume} {16}},\
  \bibinfo {pages} {509} (\bibinfo {year} {2020}{\natexlab{b}})}\BibitemShut
  {NoStop}%
\bibitem [{\citenamefont {Hillmann}\ \emph {et~al.}(2020)\citenamefont
  {Hillmann}, \citenamefont {Quijandría}, \citenamefont {Johansson},
  \citenamefont {Ferraro}, \citenamefont {Gasparinetti},\ and\ \citenamefont
  {Ferrini}}]{hillmann_universal_2020}%
  \BibitemOpen
  \bibfield  {author} {\bibinfo {author} {\bibfnamefont {T.}~\bibnamefont
  {Hillmann}}, \bibinfo {author} {\bibfnamefont {F.}~\bibnamefont
  {Quijandría}}, \bibinfo {author} {\bibfnamefont {G.}~\bibnamefont
  {Johansson}}, \bibinfo {author} {\bibfnamefont {A.}~\bibnamefont {Ferraro}},
  \bibinfo {author} {\bibfnamefont {S.}~\bibnamefont {Gasparinetti}},\ and\
  \bibinfo {author} {\bibfnamefont {G.}~\bibnamefont {Ferrini}},\ }\bibfield
  {title} {\emph {\bibinfo {title} {Universal {Gate} {Set} for
  {Continuous}-{Variable} {Quantum} {Computation} with {Microwave}
  {Circuits}}},\ }\href {https://doi.org/10.1103/PhysRevLett.125.160501}
  {\bibfield  {journal} {\bibinfo  {journal} {Phys. Rev. Lett.}\ }\textbf
  {\bibinfo {volume} {125}},\ \bibinfo {pages} {160501} (\bibinfo {year}
  {2020})}\BibitemShut {NoStop}%
\bibitem [{\citenamefont {Wustmann}\ and\ \citenamefont
  {Shumeiko}(2013)}]{wustmann_parametric_2013}%
  \BibitemOpen
  \bibfield  {author} {\bibinfo {author} {\bibfnamefont {W.}~\bibnamefont
  {Wustmann}}\ and\ \bibinfo {author} {\bibfnamefont {V.}~\bibnamefont
  {Shumeiko}},\ }\bibfield  {title} {\emph {\bibinfo {title} {Parametric
  resonance in tunable superconducting cavities}},\ }\href
  {https://doi.org/10.1103/PhysRevB.87.184501} {\bibfield  {journal} {\bibinfo
  {journal} {Phys. Rev. B}\ }\textbf {\bibinfo {volume} {87}},\ \bibinfo
  {pages} {184501} (\bibinfo {year} {2013})}\BibitemShut {NoStop}%
\bibitem [{\citenamefont {Wustmann}\ and\ \citenamefont
  {Shumeiko}(2017)}]{wustmann_nondegenerate_2017}%
  \BibitemOpen
  \bibfield  {author} {\bibinfo {author} {\bibfnamefont {W.}~\bibnamefont
  {Wustmann}}\ and\ \bibinfo {author} {\bibfnamefont {V.}~\bibnamefont
  {Shumeiko}},\ }\bibfield  {title} {\emph {\bibinfo {title} {Nondegenerate
  parametric resonance in a tunable superconducting cavity}},\ }\href
  {https://doi.org/10.1103/PhysRevApplied.8.024018} {\bibfield  {journal}
  {\bibinfo  {journal} {Phys. Rev. Applied}\ }\textbf {\bibinfo {volume} {8}},\
  \bibinfo {pages} {024018} (\bibinfo {year} {2017})}\BibitemShut {NoStop}%
\bibitem [{\citenamefont {Bergeal}\ \emph
  {et~al.}(2010{\natexlab{a}})\citenamefont {Bergeal}, \citenamefont
  {Schackert}, \citenamefont {Metcalfe}, \citenamefont {Vijay}, \citenamefont
  {Manucharyan}, \citenamefont {Frunzio}, \citenamefont {Prober}, \citenamefont
  {Schoelkopf}, \citenamefont {Girvin},\ and\ \citenamefont
  {Devoret}}]{bergeal_phase-preserving_2010}%
  \BibitemOpen
  \bibfield  {author} {\bibinfo {author} {\bibfnamefont {N.}~\bibnamefont
  {Bergeal}}, \bibinfo {author} {\bibfnamefont {F.}~\bibnamefont {Schackert}},
  \bibinfo {author} {\bibfnamefont {M.}~\bibnamefont {Metcalfe}}, \bibinfo
  {author} {\bibfnamefont {R.}~\bibnamefont {Vijay}}, \bibinfo {author}
  {\bibfnamefont {V.~E.}\ \bibnamefont {Manucharyan}}, \bibinfo {author}
  {\bibfnamefont {L.}~\bibnamefont {Frunzio}}, \bibinfo {author} {\bibfnamefont
  {D.~E.}\ \bibnamefont {Prober}}, \bibinfo {author} {\bibfnamefont {R.~J.}\
  \bibnamefont {Schoelkopf}}, \bibinfo {author} {\bibfnamefont {S.~M.}\
  \bibnamefont {Girvin}},\ and\ \bibinfo {author} {\bibfnamefont {M.~H.}\
  \bibnamefont {Devoret}},\ }\bibfield  {title} {\emph {\bibinfo {title}
  {Phase-preserving amplification near the quantum limit with a {Josephson}
  ring modulator}},\ }\href {https://doi.org/10.1038/nature09035} {\bibfield
  {journal} {\bibinfo  {journal} {Nature}\ }\textbf {\bibinfo {volume} {465}},\
  \bibinfo {pages} {64} (\bibinfo {year} {2010}{\natexlab{a}})}\BibitemShut
  {NoStop}%
\bibitem [{\citenamefont {Frattini}\ \emph {et~al.}(2018)\citenamefont
  {Frattini}, \citenamefont {Sivak}, \citenamefont {Lingenfelter},
  \citenamefont {Shankar},\ and\ \citenamefont
  {Devoret}}]{frattini_optimizing_2018}%
  \BibitemOpen
  \bibfield  {author} {\bibinfo {author} {\bibfnamefont {N.~E.}\ \bibnamefont
  {Frattini}}, \bibinfo {author} {\bibfnamefont {V.~V.}\ \bibnamefont {Sivak}},
  \bibinfo {author} {\bibfnamefont {A.}~\bibnamefont {Lingenfelter}}, \bibinfo
  {author} {\bibfnamefont {S.}~\bibnamefont {Shankar}},\ and\ \bibinfo {author}
  {\bibfnamefont {M.~H.}\ \bibnamefont {Devoret}},\ }\bibfield  {title} {\emph
  {\bibinfo {title} {Optimizing the {Nonlinearity} and {Dissipation} of a
  {SNAIL} {Parametric} {Amplifier} for {Dynamic} {Range}}},\ }\href
  {https://doi.org/10.1103/PhysRevApplied.10.054020} {\bibfield  {journal}
  {\bibinfo  {journal} {Phys. Rev. Applied}\ }\textbf {\bibinfo {volume}
  {10}},\ \bibinfo {pages} {054020} (\bibinfo {year} {2018})}\BibitemShut
  {NoStop}%
\bibitem [{\citenamefont {Sivak}\ \emph {et~al.}(2019)\citenamefont {Sivak},
  \citenamefont {Frattini}, \citenamefont {Joshi}, \citenamefont
  {Lingenfelter}, \citenamefont {Shankar},\ and\ \citenamefont
  {Devoret}}]{sivak_kerr-free_2019}%
  \BibitemOpen
  \bibfield  {author} {\bibinfo {author} {\bibfnamefont {V.}~\bibnamefont
  {Sivak}}, \bibinfo {author} {\bibfnamefont {N.}~\bibnamefont {Frattini}},
  \bibinfo {author} {\bibfnamefont {V.}~\bibnamefont {Joshi}}, \bibinfo
  {author} {\bibfnamefont {A.}~\bibnamefont {Lingenfelter}}, \bibinfo {author}
  {\bibfnamefont {S.}~\bibnamefont {Shankar}},\ and\ \bibinfo {author}
  {\bibfnamefont {M.}~\bibnamefont {Devoret}},\ }\bibfield  {title} {\emph
  {\bibinfo {title} {Kerr-{Free} {Three}-{Wave} {Mixing} in {Superconducting}
  {Quantum} {Circuits}}},\ }\href
  {https://doi.org/10.1103/PhysRevApplied.11.054060} {\bibfield  {journal}
  {\bibinfo  {journal} {Phys. Rev. Applied}\ }\textbf {\bibinfo {volume}
  {11}},\ \bibinfo {pages} {054060} (\bibinfo {year} {2019})}\BibitemShut
  {NoStop}%
\bibitem [{\citenamefont {Svensson}\ \emph {et~al.}(2018)\citenamefont
  {Svensson}, \citenamefont {Bengtsson}, \citenamefont {Bylander},
  \citenamefont {Shumeiko},\ and\ \citenamefont
  {Delsing}}]{svensson_period_2018}%
  \BibitemOpen
  \bibfield  {author} {\bibinfo {author} {\bibfnamefont {I.-M.}\ \bibnamefont
  {Svensson}}, \bibinfo {author} {\bibfnamefont {A.}~\bibnamefont {Bengtsson}},
  \bibinfo {author} {\bibfnamefont {J.}~\bibnamefont {Bylander}}, \bibinfo
  {author} {\bibfnamefont {V.}~\bibnamefont {Shumeiko}},\ and\ \bibinfo
  {author} {\bibfnamefont {P.}~\bibnamefont {Delsing}},\ }\bibfield  {title}
  {\emph {\bibinfo {title} {Period multiplication in a parametrically driven
  superconducting resonator}},\ }\href {https://doi.org/10.1063/1.5026974}
  {\bibfield  {journal} {\bibinfo  {journal} {Appl. Phys. Lett.}\ }\textbf
  {\bibinfo {volume} {113}},\ \bibinfo {pages} {022602} (\bibinfo {year}
  {2018})}\BibitemShut {NoStop}%
\bibitem [{\citenamefont {Chang}\ \emph {et~al.}(2020)\citenamefont {Chang},
  \citenamefont {Sabín}, \citenamefont {Forn-Díaz}, \citenamefont
  {Quijandría}, \citenamefont {Vadiraj}, \citenamefont {Nsanzineza},
  \citenamefont {Johansson},\ and\ \citenamefont
  {Wilson}}]{chang_observation_2020}%
  \BibitemOpen
  \bibfield  {author} {\bibinfo {author} {\bibfnamefont {C.~S.}\ \bibnamefont
  {Chang}}, \bibinfo {author} {\bibfnamefont {C.}~\bibnamefont {Sabín}},
  \bibinfo {author} {\bibfnamefont {P.}~\bibnamefont {Forn-Díaz}}, \bibinfo
  {author} {\bibfnamefont {F.}~\bibnamefont {Quijandría}}, \bibinfo {author}
  {\bibfnamefont {A.}~\bibnamefont {Vadiraj}}, \bibinfo {author} {\bibfnamefont
  {I.}~\bibnamefont {Nsanzineza}}, \bibinfo {author} {\bibfnamefont
  {G.}~\bibnamefont {Johansson}},\ and\ \bibinfo {author} {\bibfnamefont
  {C.}~\bibnamefont {Wilson}},\ }\bibfield  {title} {\emph {\bibinfo {title}
  {Observation of {Three}-{Photon} {Spontaneous} {Parametric}
  {Down}-{Conversion} in a {Superconducting} {Parametric} {Cavity}}},\ }\href
  {https://doi.org/10.1103/PhysRevX.10.011011} {\bibfield  {journal} {\bibinfo
  {journal} {Phys. Rev. X}\ }\textbf {\bibinfo {volume} {10}},\ \bibinfo
  {pages} {011011} (\bibinfo {year} {2020})}\BibitemShut {NoStop}%
\bibitem [{\citenamefont {Vool}(2017)}]{vool_engineering_2017}%
  \BibitemOpen
  \bibfield  {author} {\bibinfo {author} {\bibfnamefont {U.}~\bibnamefont
  {Vool}},\ }\emph {\bibinfo {title} {Engineering {Synthetic} {Quantum}
  {Operations}}},\ \href@noop {} {Ph.D. thesis},\ \bibinfo  {school} {Yale
  University} (\bibinfo {year} {2017})\BibitemShut {NoStop}%
\bibitem [{Note98()}]{Note98}%
  \BibitemOpen
  \bibinfo {note} {To be precise, three-wave mixing has already been achieved
  with Josephson ring modulators (JRM)~\cite {bergeal_phase-preserving_2010,
  bergeal_analog_2010}. However, the JRM has the drawback that it is a
  quadropole element and thus provides only a trilinear Hamiltonian $\varphi _X
  \varphi _Y \varphi _Z$ between three modes $X, Y$ and $Z$.}\BibitemShut
  {Stop}%
\bibitem [{\citenamefont {Miano}\ \emph {et~al.}(2021)\citenamefont {Miano},
  \citenamefont {Liu}, \citenamefont {Sivak}, \citenamefont {Frunzio},
  \citenamefont {Joshi}, \citenamefont {Dai}, \citenamefont {Frattini},\ and\
  \citenamefont {Devoret}}]{miano_full_2021}%
  \BibitemOpen
  \bibfield  {author} {\bibinfo {author} {\bibfnamefont {A.}~\bibnamefont
  {Miano}}, \bibinfo {author} {\bibfnamefont {G.}~\bibnamefont {Liu}}, \bibinfo
  {author} {\bibfnamefont {V.}~\bibnamefont {Sivak}}, \bibinfo {author}
  {\bibfnamefont {L.}~\bibnamefont {Frunzio}}, \bibinfo {author} {\bibfnamefont
  {V.}~\bibnamefont {Joshi}}, \bibinfo {author} {\bibfnamefont
  {W.}~\bibnamefont {Dai}}, \bibinfo {author} {\bibfnamefont {N.}~\bibnamefont
  {Frattini}},\ and\ \bibinfo {author} {\bibfnamefont {M.}~\bibnamefont
  {Devoret}},\ }\href {https://meetings.aps.org/Meeting/MAR21/Session/Y32.6}
  {\bibinfo {title} {Full control of {Josephson} nonlinear processes in a
  {Gradiometric} {SNAIL} {Parametric} {Amplifier}}} (\bibinfo {year} {2021}),\
  \bibinfo {note} {Bulletin of the American Physical Society 2021 -
  Y32.00006}\BibitemShut {NoStop}%
\bibitem [{\citenamefont {Grimsmo}\ \emph {et~al.}(2021)\citenamefont
  {Grimsmo}, \citenamefont {Royer}, \citenamefont {Kreikebaum}, \citenamefont
  {Ye}, \citenamefont {O’Brien}, \citenamefont {Siddiqi},\ and\ \citenamefont
  {Blais}}]{grimsmo_quantum_2021}%
  \BibitemOpen
  \bibfield  {author} {\bibinfo {author} {\bibfnamefont {A.~L.}\ \bibnamefont
  {Grimsmo}}, \bibinfo {author} {\bibfnamefont {B.}~\bibnamefont {Royer}},
  \bibinfo {author} {\bibfnamefont {J.~M.}\ \bibnamefont {Kreikebaum}},
  \bibinfo {author} {\bibfnamefont {Y.}~\bibnamefont {Ye}}, \bibinfo {author}
  {\bibfnamefont {K.}~\bibnamefont {O’Brien}}, \bibinfo {author}
  {\bibfnamefont {I.}~\bibnamefont {Siddiqi}},\ and\ \bibinfo {author}
  {\bibfnamefont {A.}~\bibnamefont {Blais}},\ }\bibfield  {title} {\emph
  {\bibinfo {title} {Quantum {Metamaterial} for {Broadband} {Detection} of
  {Single} {Microwave} {Photons}}},\ }\href
  {https://doi.org/10.1103/PhysRevApplied.15.034074} {\bibfield  {journal}
  {\bibinfo  {journal} {Phys. Rev. Applied}\ }\textbf {\bibinfo {volume}
  {15}},\ \bibinfo {pages} {034074} (\bibinfo {year} {2021})}\BibitemShut
  {NoStop}%
\bibitem [{\citenamefont {Noguchi}\ \emph
  {et~al.}(2020{\natexlab{a}})\citenamefont {Noguchi}, \citenamefont {Osada},
  \citenamefont {Masuda}, \citenamefont {Kono}, \citenamefont {Heya},
  \citenamefont {Wolski}, \citenamefont {Takahashi}, \citenamefont {Sugiyama},
  \citenamefont {Lachance-Quirion},\ and\ \citenamefont
  {Nakamura}}]{noguchi_fast_2020}%
  \BibitemOpen
  \bibfield  {author} {\bibinfo {author} {\bibfnamefont {A.}~\bibnamefont
  {Noguchi}}, \bibinfo {author} {\bibfnamefont {A.}~\bibnamefont {Osada}},
  \bibinfo {author} {\bibfnamefont {S.}~\bibnamefont {Masuda}}, \bibinfo
  {author} {\bibfnamefont {S.}~\bibnamefont {Kono}}, \bibinfo {author}
  {\bibfnamefont {K.}~\bibnamefont {Heya}}, \bibinfo {author} {\bibfnamefont
  {S.~P.}\ \bibnamefont {Wolski}}, \bibinfo {author} {\bibfnamefont
  {H.}~\bibnamefont {Takahashi}}, \bibinfo {author} {\bibfnamefont
  {T.}~\bibnamefont {Sugiyama}}, \bibinfo {author} {\bibfnamefont
  {D.}~\bibnamefont {Lachance-Quirion}},\ and\ \bibinfo {author} {\bibfnamefont
  {Y.}~\bibnamefont {Nakamura}},\ }\bibfield  {title} {\emph {\bibinfo {title}
  {Fast parametric two-qubit gates with suppressed residual interaction using
  the second-order nonlinearity of a cubic transmon}},\ }\href
  {https://doi.org/10.1103/PhysRevA.102.062408} {\bibfield  {journal} {\bibinfo
   {journal} {Phys. Rev. A}\ }\textbf {\bibinfo {volume} {102}},\ \bibinfo
  {pages} {062408} (\bibinfo {year} {2020}{\natexlab{a}})}\BibitemShut
  {NoStop}%
\bibitem [{\citenamefont {Noguchi}\ \emph
  {et~al.}(2020{\natexlab{b}})\citenamefont {Noguchi}, \citenamefont
  {Yamazaki}, \citenamefont {Tabuchi},\ and\ \citenamefont
  {Nakamura}}]{noguchi_single-photon_2020}%
  \BibitemOpen
  \bibfield  {author} {\bibinfo {author} {\bibfnamefont {A.}~\bibnamefont
  {Noguchi}}, \bibinfo {author} {\bibfnamefont {R.}~\bibnamefont {Yamazaki}},
  \bibinfo {author} {\bibfnamefont {Y.}~\bibnamefont {Tabuchi}},\ and\ \bibinfo
  {author} {\bibfnamefont {Y.}~\bibnamefont {Nakamura}},\ }\bibfield  {title}
  {\emph {\bibinfo {title} {Single-photon quantum regime of artificial
  radiation pressure on a surface acoustic wave resonator}},\ }\href
  {https://doi.org/10.1038/s41467-020-14910-z} {\bibfield  {journal} {\bibinfo
  {journal} {Nat. Commun.}\ }\textbf {\bibinfo {volume} {11}},\ \bibinfo
  {pages} {1183} (\bibinfo {year} {2020}{\natexlab{b}})}\BibitemShut {NoStop}%
\bibitem [{\citenamefont
  {Schrödinger}(1926)}]{schrodinger_quantisierung_1926}%
  \BibitemOpen
  \bibfield  {author} {\bibinfo {author} {\bibfnamefont {E.}~\bibnamefont
  {Schrödinger}},\ }\bibfield  {title} {\emph {\bibinfo {title} {Quantisierung
  als {Eigenwertproblem}}},\ }\href {https://doi.org/10.1002/andp.19263851302}
  {\bibfield  {journal} {\bibinfo  {journal} {Annalen der Physik}\ }\textbf
  {\bibinfo {volume} {385}},\ \bibinfo {pages} {437} (\bibinfo {year}
  {1926})}\BibitemShut {NoStop}%
\bibitem [{\citenamefont {James}\ and\ \citenamefont
  {Jerke}(2007)}]{james_effective_2007}%
  \BibitemOpen
  \bibfield  {author} {\bibinfo {author} {\bibfnamefont {D.~F.}\ \bibnamefont
  {James}}\ and\ \bibinfo {author} {\bibfnamefont {J.}~\bibnamefont {Jerke}},\
  }\bibfield  {title} {\emph {\bibinfo {title} {Effective {Hamiltonian} theory
  and its applications in quantum information}},\ }\href
  {https://doi.org/10.1139/p07-060} {\bibfield  {journal} {\bibinfo  {journal}
  {Can. J. Phys.}\ }\textbf {\bibinfo {volume} {85}},\ \bibinfo {pages} {625}
  (\bibinfo {year} {2007})}\BibitemShut {NoStop}%
\bibitem [{\citenamefont {Bravyi}\ \emph {et~al.}(2011)\citenamefont {Bravyi},
  \citenamefont {DiVincenzo},\ and\ \citenamefont
  {Loss}}]{bravyi_schriefferwolff_2011}%
  \BibitemOpen
  \bibfield  {author} {\bibinfo {author} {\bibfnamefont {S.}~\bibnamefont
  {Bravyi}}, \bibinfo {author} {\bibfnamefont {D.~P.}\ \bibnamefont
  {DiVincenzo}},\ and\ \bibinfo {author} {\bibfnamefont {D.}~\bibnamefont
  {Loss}},\ }\bibfield  {title} {\emph {\bibinfo {title} {Schrieffer–{Wolff}
  transformation for quantum many-body systems}},\ }\href
  {https://doi.org/10.1016/j.aop.2011.06.004} {\bibfield  {journal} {\bibinfo
  {journal} {Ann. Phys. (NY)}\ }\textbf {\bibinfo {volume} {326}},\ \bibinfo
  {pages} {2793} (\bibinfo {year} {2011})}\BibitemShut {NoStop}%
\bibitem [{\citenamefont {Paulisch}\ \emph {et~al.}(2014)\citenamefont
  {Paulisch}, \citenamefont {Rui}, \citenamefont {Ng},\ and\ \citenamefont
  {Englert}}]{paulisch_beyond_2014}%
  \BibitemOpen
  \bibfield  {author} {\bibinfo {author} {\bibfnamefont {V.}~\bibnamefont
  {Paulisch}}, \bibinfo {author} {\bibfnamefont {H.}~\bibnamefont {Rui}},
  \bibinfo {author} {\bibfnamefont {H.~K.}\ \bibnamefont {Ng}},\ and\ \bibinfo
  {author} {\bibfnamefont {B.-G.}\ \bibnamefont {Englert}},\ }\bibfield
  {title} {\emph {\bibinfo {title} {Beyond adiabatic elimination: {A} hierarchy
  of approximations for multi-photon processes}},\ }\href
  {https://doi.org/10.1140/epjp/i2014-14012-8} {\bibfield  {journal} {\bibinfo
  {journal} {Eur. Phys. J. Plus}\ }\textbf {\bibinfo {volume} {129}},\ \bibinfo
  {pages} {12} (\bibinfo {year} {2014})}\BibitemShut {NoStop}%
\bibitem [{\citenamefont {Zeuch}\ \emph {et~al.}(2020)\citenamefont {Zeuch},
  \citenamefont {Hassler}, \citenamefont {Slim},\ and\ \citenamefont
  {DiVincenzo}}]{zeuch_exact_2020}%
  \BibitemOpen
  \bibfield  {author} {\bibinfo {author} {\bibfnamefont {D.}~\bibnamefont
  {Zeuch}}, \bibinfo {author} {\bibfnamefont {F.}~\bibnamefont {Hassler}},
  \bibinfo {author} {\bibfnamefont {J.~J.}\ \bibnamefont {Slim}},\ and\
  \bibinfo {author} {\bibfnamefont {D.~P.}\ \bibnamefont {DiVincenzo}},\
  }\bibfield  {title} {\emph {\bibinfo {title} {Exact rotating wave
  approximation}},\ }\href {https://doi.org/10.1016/j.aop.2020.168327}
  {\bibfield  {journal} {\bibinfo  {journal} {Ann. Phys. (NY)}\ }\textbf
  {\bibinfo {volume} {423}},\ \bibinfo {pages} {168327} (\bibinfo {year}
  {2020})}\BibitemShut {NoStop}%
\bibitem [{\citenamefont {Schleich}(2001)}]{schleich_quantum_2001}%
  \BibitemOpen
  \bibfield  {author} {\bibinfo {author} {\bibfnamefont {W.}~\bibnamefont
  {Schleich}},\ }\href@noop {} {\emph {\bibinfo {title} {Quantum {Optics} in
  {Phase} {Space}}}}\ (\bibinfo  {publisher} {Wiley-VCH},\ \bibinfo {address}
  {Berlin; New York},\ \bibinfo {year} {2001})\BibitemShut {NoStop}%
\bibitem [{\citenamefont {James}(2000)}]{james_quantum_2000}%
  \BibitemOpen
  \bibfield  {author} {\bibinfo {author} {\bibfnamefont {D.~F.~V.}\
  \bibnamefont {James}},\ }\bibfield  {title} {\emph {\bibinfo {title} {Quantum
  {Computation} with {Hot} and {Cold} {Ions}: {An} {Assessment} of {Proposed}
  {Schemes}}},\ }\href
  {https://doi.org/10.1002/1521-3978(200009)48:9/11<823::AID-PROP823>3.0.CO;2-M}
  {\bibfield  {journal} {\bibinfo  {journal} {Fortschr. Phys.}\ }\textbf
  {\bibinfo {volume} {48}},\ \bibinfo {pages} {823} (\bibinfo {year}
  {2000})}\BibitemShut {NoStop}%
\bibitem [{\citenamefont {Gamel}\ and\ \citenamefont
  {James}(2010)}]{gamel_time-averaged_2010}%
  \BibitemOpen
  \bibfield  {author} {\bibinfo {author} {\bibfnamefont {O.}~\bibnamefont
  {Gamel}}\ and\ \bibinfo {author} {\bibfnamefont {D.~F.~V.}\ \bibnamefont
  {James}},\ }\bibfield  {title} {\emph {\bibinfo {title} {Time-averaged
  quantum dynamics and the validity of the effective {Hamiltonian} model}},\
  }\href {https://doi.org/10.1103/PhysRevA.82.052106} {\bibfield  {journal}
  {\bibinfo  {journal} {Phys. Rev. A}\ }\textbf {\bibinfo {volume} {82}},\
  \bibinfo {pages} {052106} (\bibinfo {year} {2010})}\BibitemShut {NoStop}%
\bibitem [{\citenamefont {Kessler}(2012)}]{kessler_generalized_2012}%
  \BibitemOpen
  \bibfield  {author} {\bibinfo {author} {\bibfnamefont {E.~M.}\ \bibnamefont
  {Kessler}},\ }\bibfield  {title} {\emph {\bibinfo {title} {Generalized
  {Schrieffer}-{Wolff} formalism for dissipative systems}},\ }\href
  {https://doi.org/10.1103/PhysRevA.86.012126} {\bibfield  {journal} {\bibinfo
  {journal} {Phys. Rev. A}\ }\textbf {\bibinfo {volume} {86}},\ \bibinfo
  {pages} {012126} (\bibinfo {year} {2012})}\BibitemShut {NoStop}%
\bibitem [{Note44()}]{Note44}%
  \BibitemOpen
  \bibinfo {note} {If $\protect \hat {V}$ contains multiple energy scales, it
  can be useful to introduce multiple parameters and $\lambda _i$ and split
  $\protect \hat {V}$ into multiple terms, i.e., $\lambda \protect \hat {V}
  \DOTSB \mathrel {\mathop {\kern 0pt =}\limits ^{\textstyle .}}\DOTSB \sum@
  \slimits@ _i \lambda _i \protect \hat {V}_i$.}\BibitemShut {Stop}%
\bibitem [{\citenamefont {Poletto}\ \emph {et~al.}(2012)\citenamefont
  {Poletto}, \citenamefont {Gambetta}, \citenamefont {Merkel}, \citenamefont
  {Smolin}, \citenamefont {Chow}, \citenamefont {Corcoles}, \citenamefont
  {Keefe}, \citenamefont {Rothwell}, \citenamefont {Rozen}, \citenamefont
  {Abraham}, \citenamefont {Rigetti},\ and\ \citenamefont
  {Steffen}}]{poletto_entanglement_2012}%
  \BibitemOpen
  \bibfield  {author} {\bibinfo {author} {\bibfnamefont {S.}~\bibnamefont
  {Poletto}}, \bibinfo {author} {\bibfnamefont {J.~M.}\ \bibnamefont
  {Gambetta}}, \bibinfo {author} {\bibfnamefont {S.~T.}\ \bibnamefont
  {Merkel}}, \bibinfo {author} {\bibfnamefont {J.~A.}\ \bibnamefont {Smolin}},
  \bibinfo {author} {\bibfnamefont {J.~M.}\ \bibnamefont {Chow}}, \bibinfo
  {author} {\bibfnamefont {A.~D.}\ \bibnamefont {Corcoles}}, \bibinfo {author}
  {\bibfnamefont {G.~A.}\ \bibnamefont {Keefe}}, \bibinfo {author}
  {\bibfnamefont {M.~B.}\ \bibnamefont {Rothwell}}, \bibinfo {author}
  {\bibfnamefont {J.~R.}\ \bibnamefont {Rozen}}, \bibinfo {author}
  {\bibfnamefont {D.~W.}\ \bibnamefont {Abraham}}, \bibinfo {author}
  {\bibfnamefont {C.}~\bibnamefont {Rigetti}},\ and\ \bibinfo {author}
  {\bibfnamefont {M.}~\bibnamefont {Steffen}},\ }\bibfield  {title} {\emph
  {\bibinfo {title} {Entanglement of two superconducting qubits in a waveguide
  cavity via monochromatic two-photon excitation}},\ }\href
  {https://doi.org/10.1103/PhysRevLett.109.240505} {\bibfield  {journal}
  {\bibinfo  {journal} {Phys. Rev. Lett.}\ }\textbf {\bibinfo {volume} {109}},\
  \bibinfo {pages} {240505} (\bibinfo {year} {2012})}\BibitemShut {NoStop}%
\bibitem [{\citenamefont {Winkler}(2003)}]{winkler_spin_2003}%
  \BibitemOpen
  \bibfield  {author} {\bibinfo {author} {\bibfnamefont {R.}~\bibnamefont
  {Winkler}},\ }\href {https://doi.org/10.1007/b13586} {\emph {\bibinfo {title}
  {Spin{\textemdash}Orbit Coupling Effects in Two-Dimensional Electron and Hole
  Systems}}}\ (\bibinfo  {publisher} {Springer Berlin Heidelberg},\ \bibinfo
  {year} {2003})\BibitemShut {NoStop}%
\bibitem [{\citenamefont {Magesan}\ and\ \citenamefont
  {Gambetta}(2020)}]{magesan_effective_2020}%
  \BibitemOpen
  \bibfield  {author} {\bibinfo {author} {\bibfnamefont {E.}~\bibnamefont
  {Magesan}}\ and\ \bibinfo {author} {\bibfnamefont {J.~M.}\ \bibnamefont
  {Gambetta}},\ }\bibfield  {title} {\emph {\bibinfo {title} {Effective
  {Hamiltonian} models of the cross-resonance gate}},\ }\href
  {https://doi.org/10.1103/PhysRevA.101.052308} {\bibfield  {journal} {\bibinfo
   {journal} {Phys. Rev. A}\ }\textbf {\bibinfo {volume} {101}},\ \bibinfo
  {pages} {052308} (\bibinfo {year} {2020})}\BibitemShut {NoStop}%
\bibitem [{\citenamefont {Bukov}\ \emph {et~al.}(2015)\citenamefont {Bukov},
  \citenamefont {D'Alessio},\ and\ \citenamefont
  {Polkovnikov}}]{bukov_universal_2015}%
  \BibitemOpen
  \bibfield  {author} {\bibinfo {author} {\bibfnamefont {M.}~\bibnamefont
  {Bukov}}, \bibinfo {author} {\bibfnamefont {L.}~\bibnamefont {D'Alessio}},\
  and\ \bibinfo {author} {\bibfnamefont {A.}~\bibnamefont {Polkovnikov}},\
  }\bibfield  {title} {\emph {\bibinfo {title} {Universal high-frequency
  behavior of periodically driven systems: from dynamical stabilization to
  floquet engineering}},\ }\href
  {https://doi.org/10.1080/00018732.2015.1055918} {\bibfield  {journal}
  {\bibinfo  {journal} {Advances in Physics}\ }\textbf {\bibinfo {volume}
  {64}},\ \bibinfo {pages} {139} (\bibinfo {year} {2015})},\ \Eprint
  {https://arxiv.org/abs/https://doi.org/10.1080/00018732.2015.1055918}
  {https://doi.org/10.1080/00018732.2015.1055918} \BibitemShut {NoStop}%
\bibitem [{\citenamefont {Zinn-Justin}(2002)}]{zinn-justin_quantum_2002}%
  \BibitemOpen
  \bibfield  {author} {\bibinfo {author} {\bibfnamefont {J.}~\bibnamefont
  {Zinn-Justin}},\ }\href
  {https://doi.org/10.1093/acprof:oso/9780198509233.001.0001} {\emph {\bibinfo
  {title} {Quantum {Field} {Theory} and {Critical} {Phenomena}}}}\ (\bibinfo
  {publisher} {Oxford University Press},\ \bibinfo {year} {2002})\BibitemShut
  {NoStop}%
\bibitem [{\citenamefont {Axline}\ \emph {et~al.}(2018)\citenamefont {Axline},
  \citenamefont {Burkhart}, \citenamefont {Pfaff}, \citenamefont {Zhang},
  \citenamefont {Chou}, \citenamefont {Campagne-Ibarcq}, \citenamefont
  {Reinhold}, \citenamefont {Frunzio}, \citenamefont {Girvin}, \citenamefont
  {Jiang}, \citenamefont {Devoret},\ and\ \citenamefont
  {Schoelkopf}}]{axline_-demand_2018}%
  \BibitemOpen
  \bibfield  {author} {\bibinfo {author} {\bibfnamefont {C.~J.}\ \bibnamefont
  {Axline}}, \bibinfo {author} {\bibfnamefont {L.~D.}\ \bibnamefont
  {Burkhart}}, \bibinfo {author} {\bibfnamefont {W.}~\bibnamefont {Pfaff}},
  \bibinfo {author} {\bibfnamefont {M.}~\bibnamefont {Zhang}}, \bibinfo
  {author} {\bibfnamefont {K.}~\bibnamefont {Chou}}, \bibinfo {author}
  {\bibfnamefont {P.}~\bibnamefont {Campagne-Ibarcq}}, \bibinfo {author}
  {\bibfnamefont {P.}~\bibnamefont {Reinhold}}, \bibinfo {author}
  {\bibfnamefont {L.}~\bibnamefont {Frunzio}}, \bibinfo {author} {\bibfnamefont
  {S.~M.}\ \bibnamefont {Girvin}}, \bibinfo {author} {\bibfnamefont
  {L.}~\bibnamefont {Jiang}}, \bibinfo {author} {\bibfnamefont {M.~H.}\
  \bibnamefont {Devoret}},\ and\ \bibinfo {author} {\bibfnamefont {R.~J.}\
  \bibnamefont {Schoelkopf}},\ }\bibfield  {title} {\emph {\bibinfo {title}
  {On-demand quantum state transfer and entanglement between remote microwave
  cavity memories}},\ }\href {https://doi.org/10.1038/s41567-018-0115-y}
  {\bibfield  {journal} {\bibinfo  {journal} {Nat. Phys.}\ }\textbf {\bibinfo
  {volume} {14}},\ \bibinfo {pages} {705} (\bibinfo {year} {2018})}\BibitemShut
  {NoStop}%
\bibitem [{\citenamefont {Burkhart}\ \emph {et~al.}(2020)\citenamefont
  {Burkhart}, \citenamefont {Teoh}, \citenamefont {Zhang}, \citenamefont
  {Axline}, \citenamefont {Frunzio}, \citenamefont {Devoret}, \citenamefont
  {Jiang}, \citenamefont {Girvin},\ and\ \citenamefont
  {Schoelkopf}}]{burkhart_error-detected_2020}%
  \BibitemOpen
  \bibfield  {author} {\bibinfo {author} {\bibfnamefont {L.~D.}\ \bibnamefont
  {Burkhart}}, \bibinfo {author} {\bibfnamefont {J.}~\bibnamefont {Teoh}},
  \bibinfo {author} {\bibfnamefont {Y.}~\bibnamefont {Zhang}}, \bibinfo
  {author} {\bibfnamefont {C.~J.}\ \bibnamefont {Axline}}, \bibinfo {author}
  {\bibfnamefont {L.}~\bibnamefont {Frunzio}}, \bibinfo {author} {\bibfnamefont
  {M.~H.}\ \bibnamefont {Devoret}}, \bibinfo {author} {\bibfnamefont
  {L.}~\bibnamefont {Jiang}}, \bibinfo {author} {\bibfnamefont {S.~M.}\
  \bibnamefont {Girvin}},\ and\ \bibinfo {author} {\bibfnamefont {R.~J.}\
  \bibnamefont {Schoelkopf}},\ }\bibfield  {title} {\emph {\bibinfo {title}
  {Error-detected state transfer and entanglement in a superconducting quantum
  network}},\ }\href {http://arxiv.org/abs/2004.06168} {\bibfield  {journal}
  {\bibinfo  {journal} {arXiv:2004.06168 [quant-ph]}\ } (\bibinfo {year}
  {2020})}\BibitemShut {NoStop}%
\bibitem [{\citenamefont {Petrescu}\ \emph
  {et~al.}(2021{\natexlab{b}})\citenamefont {Petrescu}, \citenamefont {Royer},\
  and\ \citenamefont {Blais}}]{petrescu_accurate_2021}%
  \BibitemOpen
  \bibfield  {author} {\bibinfo {author} {\bibfnamefont {A.}~\bibnamefont
  {Petrescu}}, \bibinfo {author} {\bibfnamefont {B.}~\bibnamefont {Royer}},\
  and\ \bibinfo {author} {\bibfnamefont {A.}~\bibnamefont {Blais}},\ }\href
  {http://meetings.aps.org/Meeting/MAR21/Session/Y32.13} {\bibinfo {title}
  {Accurate theory for drive-activated nonlinear processes in the {SNAIL}
  parametric amplifier}} (\bibinfo {year} {2021}{\natexlab{b}}),\ \bibinfo
  {note} {Bulletin of the American Physical Society 2021 -
  Y32.00013}\BibitemShut {NoStop}%
\bibitem [{\citenamefont {Wang}\ \emph {et~al.}(2020)\citenamefont {Wang},
  \citenamefont {Curtis}, \citenamefont {Lester}, \citenamefont {Zhang},
  \citenamefont {Gao}, \citenamefont {Freeze}, \citenamefont {Batista},
  \citenamefont {Vaccaro}, \citenamefont {Chuang}, \citenamefont {Frunzio},
  \citenamefont {Jiang}, \citenamefont {Girvin},\ and\ \citenamefont
  {Schoelkopf}}]{wang_efficient_2020}%
  \BibitemOpen
  \bibfield  {author} {\bibinfo {author} {\bibfnamefont {C.~S.}\ \bibnamefont
  {Wang}}, \bibinfo {author} {\bibfnamefont {J.~C.}\ \bibnamefont {Curtis}},
  \bibinfo {author} {\bibfnamefont {B.~J.}\ \bibnamefont {Lester}}, \bibinfo
  {author} {\bibfnamefont {Y.}~\bibnamefont {Zhang}}, \bibinfo {author}
  {\bibfnamefont {Y.~Y.}\ \bibnamefont {Gao}}, \bibinfo {author} {\bibfnamefont
  {J.}~\bibnamefont {Freeze}}, \bibinfo {author} {\bibfnamefont {V.~S.}\
  \bibnamefont {Batista}}, \bibinfo {author} {\bibfnamefont {P.~H.}\
  \bibnamefont {Vaccaro}}, \bibinfo {author} {\bibfnamefont {I.~L.}\
  \bibnamefont {Chuang}}, \bibinfo {author} {\bibfnamefont {L.}~\bibnamefont
  {Frunzio}}, \bibinfo {author} {\bibfnamefont {L.}~\bibnamefont {Jiang}},
  \bibinfo {author} {\bibfnamefont {S.}~\bibnamefont {Girvin}},\ and\ \bibinfo
  {author} {\bibfnamefont {R.~J.}\ \bibnamefont {Schoelkopf}},\ }\bibfield
  {title} {\emph {\bibinfo {title} {Efficient {Multiphoton} {Sampling} of
  {Molecular} {Vibronic} {Spectra} on a {Superconducting} {Bosonic}
  {Processor}}},\ }\href {https://doi.org/10.1103/PhysRevX.10.021060}
  {\bibfield  {journal} {\bibinfo  {journal} {Phys. Rev. X}\ }\textbf {\bibinfo
  {volume} {10}},\ \bibinfo {pages} {021060} (\bibinfo {year}
  {2020})}\BibitemShut {NoStop}%
\bibitem [{\citenamefont {Yurke}\ and\ \citenamefont
  {Stoler}(1988)}]{yurke_dynamic_1988}%
  \BibitemOpen
  \bibfield  {author} {\bibinfo {author} {\bibfnamefont {B.}~\bibnamefont
  {Yurke}}\ and\ \bibinfo {author} {\bibfnamefont {D.}~\bibnamefont {Stoler}},\
  }\bibfield  {title} {\emph {\bibinfo {title} {The dynamic generation of
  {Schrödinger} cats and their detection}},\ }\href
  {https://doi.org/10.1016/0378-4363(88)90181-7} {\bibfield  {journal}
  {\bibinfo  {journal} {Physica B+C}\ }\textbf {\bibinfo {volume} {151}},\
  \bibinfo {pages} {298} (\bibinfo {year} {1988})}\BibitemShut {NoStop}%
\bibitem [{\citenamefont {Kirchmair}\ \emph {et~al.}(2013)\citenamefont
  {Kirchmair}, \citenamefont {Vlastakis}, \citenamefont {Leghtas},
  \citenamefont {Nigg}, \citenamefont {Paik}, \citenamefont {Ginossar},
  \citenamefont {Mirrahimi}, \citenamefont {Frunzio}, \citenamefont {Girvin},\
  and\ \citenamefont {Schoelkopf}}]{kirchmair_observation_2013}%
  \BibitemOpen
  \bibfield  {author} {\bibinfo {author} {\bibfnamefont {G.}~\bibnamefont
  {Kirchmair}}, \bibinfo {author} {\bibfnamefont {B.}~\bibnamefont
  {Vlastakis}}, \bibinfo {author} {\bibfnamefont {Z.}~\bibnamefont {Leghtas}},
  \bibinfo {author} {\bibfnamefont {S.~E.}\ \bibnamefont {Nigg}}, \bibinfo
  {author} {\bibfnamefont {H.}~\bibnamefont {Paik}}, \bibinfo {author}
  {\bibfnamefont {E.}~\bibnamefont {Ginossar}}, \bibinfo {author}
  {\bibfnamefont {M.}~\bibnamefont {Mirrahimi}}, \bibinfo {author}
  {\bibfnamefont {L.}~\bibnamefont {Frunzio}}, \bibinfo {author} {\bibfnamefont
  {S.~M.}\ \bibnamefont {Girvin}},\ and\ \bibinfo {author} {\bibfnamefont
  {R.~J.}\ \bibnamefont {Schoelkopf}},\ }\bibfield  {title} {\emph {\bibinfo
  {title} {Observation of quantum state collapse and revival due to the
  single-photon {Kerr} effect}},\ }\href {https://doi.org/10.1038/nature11902}
  {\bibfield  {journal} {\bibinfo  {journal} {Nature}\ }\textbf {\bibinfo
  {volume} {495}},\ \bibinfo {pages} {205} (\bibinfo {year}
  {2013})}\BibitemShut {NoStop}%
\bibitem [{\citenamefont {Elliott}\ \emph {et~al.}(2018)\citenamefont
  {Elliott}, \citenamefont {Joo},\ and\ \citenamefont
  {Ginossar}}]{elliott_designing_2018}%
  \BibitemOpen
  \bibfield  {author} {\bibinfo {author} {\bibfnamefont {M.}~\bibnamefont
  {Elliott}}, \bibinfo {author} {\bibfnamefont {J.}~\bibnamefont {Joo}},\ and\
  \bibinfo {author} {\bibfnamefont {E.}~\bibnamefont {Ginossar}},\ }\bibfield
  {title} {\emph {\bibinfo {title} {Designing {Kerr} interactions using
  multiple superconducting qubit types in a single circuit}},\ }\href
  {https://doi.org/10.1088/1367-2630/aa9243} {\bibfield  {journal} {\bibinfo
  {journal} {New J. Phys.}\ }\textbf {\bibinfo {volume} {20}},\ \bibinfo
  {pages} {023037} (\bibinfo {year} {2018})}\BibitemShut {NoStop}%
\bibitem [{\citenamefont {Gottesman}\ \emph {et~al.}(2001)\citenamefont
  {Gottesman}, \citenamefont {Kitaev},\ and\ \citenamefont
  {Preskill}}]{gottesman_encoding_2001}%
  \BibitemOpen
  \bibfield  {author} {\bibinfo {author} {\bibfnamefont {D.}~\bibnamefont
  {Gottesman}}, \bibinfo {author} {\bibfnamefont {A.}~\bibnamefont {Kitaev}},\
  and\ \bibinfo {author} {\bibfnamefont {J.}~\bibnamefont {Preskill}},\
  }\bibfield  {title} {\emph {\bibinfo {title} {Encoding a qubit in an
  oscillator}},\ }\href {https://doi.org/10.1103/PhysRevA.64.012310} {\bibfield
   {journal} {\bibinfo  {journal} {Phys. Rev. A}\ }\textbf {\bibinfo {volume}
  {64}},\ \bibinfo {pages} {012310} (\bibinfo {year} {2001})}\BibitemShut
  {NoStop}%
\bibitem [{\citenamefont {Konno}\ \emph {et~al.}(2021)\citenamefont {Konno},
  \citenamefont {Asavanant}, \citenamefont {Fukui}, \citenamefont {Sakaguchi},
  \citenamefont {Hanamura}, \citenamefont {Marek}, \citenamefont {Filip},
  \citenamefont {Yoshikawa},\ and\ \citenamefont
  {Furusawa}}]{konno_non-clifford_2021}%
  \BibitemOpen
  \bibfield  {author} {\bibinfo {author} {\bibfnamefont {S.}~\bibnamefont
  {Konno}}, \bibinfo {author} {\bibfnamefont {W.}~\bibnamefont {Asavanant}},
  \bibinfo {author} {\bibfnamefont {K.}~\bibnamefont {Fukui}}, \bibinfo
  {author} {\bibfnamefont {A.}~\bibnamefont {Sakaguchi}}, \bibinfo {author}
  {\bibfnamefont {F.}~\bibnamefont {Hanamura}}, \bibinfo {author}
  {\bibfnamefont {P.}~\bibnamefont {Marek}}, \bibinfo {author} {\bibfnamefont
  {R.}~\bibnamefont {Filip}}, \bibinfo {author} {\bibfnamefont {J.-i.}\
  \bibnamefont {Yoshikawa}},\ and\ \bibinfo {author} {\bibfnamefont
  {A.}~\bibnamefont {Furusawa}},\ }\bibfield  {title} {\emph {\bibinfo {title}
  {Non-{Clifford} gate on optical qubits by nonlinear feedforward}},\ }\href
  {http://arxiv.org/abs/2103.10644} {\bibfield  {journal} {\bibinfo  {journal}
  {arXiv:2103.10644 [quant-ph]}\ } (\bibinfo {year} {2021})}\BibitemShut
  {NoStop}%
\bibitem [{\citenamefont {Rosenblum}\ \emph {et~al.}(2018)\citenamefont
  {Rosenblum}, \citenamefont {Reinhold}, \citenamefont {Mirrahimi},
  \citenamefont {Jiang}, \citenamefont {Frunzio},\ and\ \citenamefont
  {Schoelkopf}}]{rosenblum_fault-tolerant_2018}%
  \BibitemOpen
  \bibfield  {author} {\bibinfo {author} {\bibfnamefont {S.}~\bibnamefont
  {Rosenblum}}, \bibinfo {author} {\bibfnamefont {P.}~\bibnamefont {Reinhold}},
  \bibinfo {author} {\bibfnamefont {M.}~\bibnamefont {Mirrahimi}}, \bibinfo
  {author} {\bibfnamefont {L.}~\bibnamefont {Jiang}}, \bibinfo {author}
  {\bibfnamefont {L.}~\bibnamefont {Frunzio}},\ and\ \bibinfo {author}
  {\bibfnamefont {R.~J.}\ \bibnamefont {Schoelkopf}},\ }\bibfield  {title}
  {\emph {\bibinfo {title} {Fault-tolerant detection of a quantum error}},\
  }\href {https://doi.org/10.1126/science.aat3996} {\bibfield  {journal}
  {\bibinfo  {journal} {Science}\ }\textbf {\bibinfo {volume} {361}},\ \bibinfo
  {pages} {266} (\bibinfo {year} {2018})}\BibitemShut {NoStop}%
\bibitem [{\citenamefont {Yan}\ \emph {et~al.}(2018)\citenamefont {Yan},
  \citenamefont {Krantz}, \citenamefont {Sung}, \citenamefont {Kjaergaard},
  \citenamefont {Campbell}, \citenamefont {Orlando}, \citenamefont
  {Gustavsson},\ and\ \citenamefont {Oliver}}]{yan_tunable_2018}%
  \BibitemOpen
  \bibfield  {author} {\bibinfo {author} {\bibfnamefont {F.}~\bibnamefont
  {Yan}}, \bibinfo {author} {\bibfnamefont {P.}~\bibnamefont {Krantz}},
  \bibinfo {author} {\bibfnamefont {Y.}~\bibnamefont {Sung}}, \bibinfo {author}
  {\bibfnamefont {M.}~\bibnamefont {Kjaergaard}}, \bibinfo {author}
  {\bibfnamefont {D.~L.}\ \bibnamefont {Campbell}}, \bibinfo {author}
  {\bibfnamefont {T.~P.}\ \bibnamefont {Orlando}}, \bibinfo {author}
  {\bibfnamefont {S.}~\bibnamefont {Gustavsson}},\ and\ \bibinfo {author}
  {\bibfnamefont {W.~D.}\ \bibnamefont {Oliver}},\ }\bibfield  {title} {\emph
  {\bibinfo {title} {Tunable {Coupling} {Scheme} for {Implementing}
  {High}-{Fidelity} {Two}-{Qubit} {Gates}}},\ }\href
  {https://link.aps.org/doi/10.1103/PhysRevApplied.10.054062} {\bibfield
  {journal} {\bibinfo  {journal} {Phys. Rev. Applied}\ }\textbf {\bibinfo
  {volume} {10}},\ \bibinfo {pages} {054062} (\bibinfo {year}
  {2018})}\BibitemShut {NoStop}%
\bibitem [{\citenamefont {Petrescu}\ \emph {et~al.}(2020)\citenamefont
  {Petrescu}, \citenamefont {Malekakhlagh},\ and\ \citenamefont
  {Türeci}}]{petrescu_lifetime_2020}%
  \BibitemOpen
  \bibfield  {author} {\bibinfo {author} {\bibfnamefont {A.}~\bibnamefont
  {Petrescu}}, \bibinfo {author} {\bibfnamefont {M.}~\bibnamefont
  {Malekakhlagh}},\ and\ \bibinfo {author} {\bibfnamefont {H.~E.}\ \bibnamefont
  {Türeci}},\ }\bibfield  {title} {\emph {\bibinfo {title} {Lifetime
  renormalization of driven weakly anharmonic superconducting qubits. {II}.
  {The} readout problem}},\ }\href
  {https://doi.org/10.1103/PhysRevB.101.134510} {\bibfield  {journal} {\bibinfo
   {journal} {Phys. Rev. B}\ }\textbf {\bibinfo {volume} {101}},\ \bibinfo
  {pages} {134510} (\bibinfo {year} {2020})}\BibitemShut {NoStop}%
\bibitem [{Note1()}]{Note1}%
  \BibitemOpen
  \bibinfo {note} {After completing this work, we noticed that the authors of
  Ref.~\cite {sigmund_u-matrix_1974} found a different method to calculate
  $\protect \hat S^{(1)}$. Nevertheless, to the best of our knowledge their
  approach has not been generalized to an arbitrary order in the
  perturbation.}\BibitemShut {Stop}%
\bibitem [{Note2()}]{Note2}%
  \BibitemOpen
  \bibinfo {note} {Maple 2021. Maplesoft, a division of Waterloo Maple Inc.,
  Waterloo, Ontario.}\BibitemShut {Stop}%
\bibitem [{Note3()}]{Note3}%
  \BibitemOpen
  \bibinfo {note} {The code is available at \protect \url
  {https://doi.org/10.5281/zenodo.5091424}.}\BibitemShut {Stop}%
\bibitem [{\citenamefont {Wick}(1950)}]{wick_evaluation_1950}%
  \BibitemOpen
  \bibfield  {author} {\bibinfo {author} {\bibfnamefont {G.~C.}\ \bibnamefont
  {Wick}},\ }\bibfield  {title} {\emph {\bibinfo {title} {The {Evaluation} of
  the {Collision} {Matrix}}},\ }\href {https://doi.org/10.1103/PhysRev.80.268}
  {\bibfield  {journal} {\bibinfo  {journal} {Phys. Rev.}\ }\textbf {\bibinfo
  {volume} {80}},\ \bibinfo {pages} {268} (\bibinfo {year} {1950})}\BibitemShut
  {NoStop}%
\bibitem [{\citenamefont {Blasiak}\ \emph {et~al.}(2007)\citenamefont
  {Blasiak}, \citenamefont {Horzela}, \citenamefont {Penson}, \citenamefont
  {Solomon},\ and\ \citenamefont {Duchamp}}]{blasiak_combinatorics_2007}%
  \BibitemOpen
  \bibfield  {author} {\bibinfo {author} {\bibfnamefont {P.}~\bibnamefont
  {Blasiak}}, \bibinfo {author} {\bibfnamefont {A.}~\bibnamefont {Horzela}},
  \bibinfo {author} {\bibfnamefont {K.~A.}\ \bibnamefont {Penson}}, \bibinfo
  {author} {\bibfnamefont {A.~I.}\ \bibnamefont {Solomon}},\ and\ \bibinfo
  {author} {\bibfnamefont {G.~H.~E.}\ \bibnamefont {Duchamp}},\ }\bibfield
  {title} {\emph {\bibinfo {title} {Combinatorics and {Boson} normal ordering:
  {A} gentle introduction}},\ }\href {https://doi.org/10.1119/1.2723799}
  {\bibfield  {journal} {\bibinfo  {journal} {Am. J. Phys.}\ }\textbf {\bibinfo
  {volume} {75}},\ \bibinfo {pages} {639} (\bibinfo {year} {2007})}\BibitemShut
  {NoStop}%
\bibitem [{\citenamefont {Savitzky}\ and\ \citenamefont
  {Golay}(1964)}]{savitzky_smoothing_1964}%
  \BibitemOpen
  \bibfield  {author} {\bibinfo {author} {\bibfnamefont {A.}~\bibnamefont
  {Savitzky}}\ and\ \bibinfo {author} {\bibfnamefont {M.~J.~E.}\ \bibnamefont
  {Golay}},\ }\bibfield  {title} {\emph {\bibinfo {title} {Smoothing and
  {Differentiation} of {Data} by {Simplified} {Least} {Squares}
  {Procedures}.}},\ }\href {https://doi.org/10.1021/ac60214a047} {\bibfield
  {journal} {\bibinfo  {journal} {Anal. Chem.}\ }\textbf {\bibinfo {volume}
  {36}},\ \bibinfo {pages} {1627} (\bibinfo {year} {1964})}\BibitemShut
  {NoStop}%
\bibitem [{\citenamefont {Virtanen}\ \emph {et~al.}(2020)\citenamefont
  {Virtanen} \emph {et~al.}}]{virtanen_scipy_2020}%
  \BibitemOpen
  \bibfield  {author} {\bibinfo {author} {\bibfnamefont {P.}~\bibnamefont
  {Virtanen}} \emph {et~al.},\ }\bibfield  {title} {\emph {\bibinfo {title}
  {{SciPy} 1.0: fundamental algorithms for scientific computing in {Python}}},\
  }\href {https://doi.org/10.1038/s41592-019-0686-2} {\bibfield  {journal}
  {\bibinfo  {journal} {Nat. Methods}\ }\textbf {\bibinfo {volume} {17}},\
  \bibinfo {pages} {261} (\bibinfo {year} {2020})}\BibitemShut {NoStop}%
\bibitem [{\citenamefont {Johansson}\ \emph {et~al.}(2013)\citenamefont
  {Johansson}, \citenamefont {Nation},\ and\ \citenamefont
  {Nori}}]{johansson_qutip_2013}%
  \BibitemOpen
  \bibfield  {author} {\bibinfo {author} {\bibfnamefont {J.~R.}\ \bibnamefont
  {Johansson}}, \bibinfo {author} {\bibfnamefont {P.~D.}\ \bibnamefont
  {Nation}},\ and\ \bibinfo {author} {\bibfnamefont {F.}~\bibnamefont {Nori}},\
  }\bibfield  {title} {\emph {\bibinfo {title} {{QuTiP} 2: {A} {Python}
  framework for the dynamics of open quantum systems}},\ }\href
  {https://doi.org/10.1016/j.cpc.2012.11.019} {\bibfield  {journal} {\bibinfo
  {journal} {Comput. Phys. Commun.}\ }\textbf {\bibinfo {volume} {184}},\
  \bibinfo {pages} {1234} (\bibinfo {year} {2013})}\BibitemShut {NoStop}%
\bibitem [{\citenamefont {Bergeal}\ \emph
  {et~al.}(2010{\natexlab{b}})\citenamefont {Bergeal}, \citenamefont {Vijay},
  \citenamefont {Manucharyan}, \citenamefont {Siddiqi}, \citenamefont
  {Schoelkopf}, \citenamefont {Girvin},\ and\ \citenamefont
  {Devoret}}]{bergeal_analog_2010}%
  \BibitemOpen
  \bibfield  {author} {\bibinfo {author} {\bibfnamefont {N.}~\bibnamefont
  {Bergeal}}, \bibinfo {author} {\bibfnamefont {R.}~\bibnamefont {Vijay}},
  \bibinfo {author} {\bibfnamefont {V.~E.}\ \bibnamefont {Manucharyan}},
  \bibinfo {author} {\bibfnamefont {I.}~\bibnamefont {Siddiqi}}, \bibinfo
  {author} {\bibfnamefont {R.~J.}\ \bibnamefont {Schoelkopf}}, \bibinfo
  {author} {\bibfnamefont {S.~M.}\ \bibnamefont {Girvin}},\ and\ \bibinfo
  {author} {\bibfnamefont {M.~H.}\ \bibnamefont {Devoret}},\ }\bibfield
  {title} {\emph {\bibinfo {title} {Analog information processing at the
  quantum limit with a {Josephson} ring modulator}},\ }\href
  {https://doi.org/10.1038/nphys1516} {\bibfield  {journal} {\bibinfo
  {journal} {Nat. Phys.}\ }\textbf {\bibinfo {volume} {6}},\ \bibinfo {pages}
  {296} (\bibinfo {year} {2010}{\natexlab{b}})}\BibitemShut {NoStop}%
\bibitem [{\citenamefont {Sigmund}\ and\ \citenamefont
  {Wagner}(1974)}]{sigmund_u-matrix_1974}%
  \BibitemOpen
  \bibfield  {author} {\bibinfo {author} {\bibfnamefont {E.}~\bibnamefont
  {Sigmund}}\ and\ \bibinfo {author} {\bibfnamefont {M.}~\bibnamefont
  {Wagner}},\ }\bibfield  {title} {\emph {\bibinfo {title} {The {U}-matrix as a
  simple tool to derive canonical transformations}},\ }\href
  {https://doi.org/10.1007/BF01669887} {\bibfield  {journal} {\bibinfo
  {journal} {Zeitschrift für Physik}\ }\textbf {\bibinfo {volume} {268}},\
  \bibinfo {pages} {245} (\bibinfo {year} {1974})}\BibitemShut {NoStop}%
\end{thebibliography}

\end{document}